\documentclass[aps,prl,twocolumn,10pt,superscriptaddress]{revtex4-1}

\usepackage{graphicx,color}
\usepackage{amsmath,amssymb}
\usepackage{braket,slashed}
\usepackage{hyperref}
\usepackage{subfigure}
\usepackage[english]{babel}
\usepackage[mathscr]{euscript}
\usepackage{bm}
\usepackage{textcomp}

\newcommand{\ZZ}{\ensuremath{\mathbb{Z}}}

\renewcommand{\d}{\ensuremath{\mathrm{d}}}
\newcommand{\e}{\ensuremath{\mathrm{e}}}
\newcommand{\ham}{\ensuremath{\hat{H}}}
\renewcommand{\vec}[1]{\ensuremath{\mathbf{#1}}}
\newcommand{\lv}{\ensuremath{\vec{v_L^\dagger}}}
\newcommand{\rv}{\ensuremath{\vec{v_R}}}
\newcommand{\spst}{\ensuremath{\ket{\{s\}}}}
\newcommand{\perm}{\ensuremath{{\mathcal{P}}}}
\newcommand{\ie}{\textit{i.e. }}
\newcommand{\eg}{\textit{e.g. }}
\newcommand{\tussentitel}[1]{\par\textit{#1}}

\date{\today}

\begin{document}

\title{S matrix from matrix product states}

\author{Laurens Vanderstraeten}
\author{Jutho Haegeman}
\affiliation{Ghent University, Department of Physics and Astronomy, Krijgslaan 281-S9, B-9000 Gent, Belgium}
\author{Tobias J.~Osborne}
\affiliation{Leibniz Universit\"{a}t Hannover, Institute of Theoretical Physics, Appelstrasse 2, D-30167 Hannover, Germany}
\author{Frank Verstraete}
\affiliation{Ghent University, Department of Physics and Astronomy, Krijgslaan 281-S9, B-9000 Gent, Belgium}
\affiliation{Vienna Center for Quantum Science, Universit\"at Wien, Boltzmanngasse 5, A-1090 Wien, Austria}

\begin{abstract}
We use the matrix product state formalism to construct stationary scattering states of elementary excitations in generic one-dimensional quantum lattice systems. Our method is applied to the spin-1 Heisenberg antiferromagnet, for which we calculate the full magnon-magnon S matrix for arbitrary momenta and spin, the two-particle contribution to the spectral function and the magnetization curve. As our method provides an accurate microscopic representation of the interaction between elementary excitations, we envisage the description of low-energy dynamics of one-dimensional spin chains in terms of these particlelike excitations.
\end{abstract}

\maketitle

\addcontentsline{toc}{section}{Introduction}
Theoretical studies have shown that, despite the exponential growth of Hilbert space, the low-energy physics of large one-dimensional quantum systems can be described efficiently. More specifically, the entanglement of the ground state and lowest-lying excitations obey an area law \cite{Hastings2007, *Masanes2009}, which confines the low-lying physics of these systems to some small subspace of the complete Hilbert space. A natural and efficient parametrization of this subspace is provided by the class of matrix product states (MPS) \cite{Schollwock2011a} underlying the density matrix renormalization group (DMRG) \cite{White1992}. While DMRG and MPS algorithms were initially focused on describing ground states, a lot of work has gone into extending the formalism to the calculation of dynamical properties \cite{Hallberg1995, *Kuhner1999, *Jeckelmann2002, *Verstraete2004c, *Vidal2004a, *White2004, *Weichselbaum2009a, *Holzner2011, *Dargel2012, White2008a}.
\par One of the approaches towards the low-lying dynamics consists in finding accurate descriptions of elementary excitations variationally. By casting the Feynman-Bijl ansatz \cite{Feynman1954, *Feynman1956, *Bijl1941} into the MPS formalism, elementary excitation spectra of one-dimensional quantum spin systems \cite{Haegeman2012a} and quantum field theories \cite{Draxler2013, *Milsted2013, *Buyens2013} were obtained to an unprecedented precision. Recently, this approach has found its theoretical ground as it was shown that gapped elementary excitations are local in the sense that they can be created by a local operator acting on the ground state \cite{Haegeman2013a}. This result suggested that elementary excitations can be identified as particles on a nontrivial background and raises the question whether we can study  their scattering. As the particle interactions are partly constituted by the strongly correlated background itself, this amounts to a highly nontrivial scattering problem (in contrast to Ref. \cite{Shastry1981}, where scattering states were constructed on top of a product dimer state).
\par In many studies, interactions between elementary excitations are modeled by different effective field theories to capture, for example, the response of magnetic systems to external fields \cite{Affleck1990, *Affleck1991,Tsvelik1990,Chitra1997,Konik2002}. Lacking a microscopic description of the particles, the parameters in these effective theories had to be determined from global properties of the system \cite{Lou2000,Okunishi1999} and/or strong- or weak-coupling limits \cite{Damle1998}.
\par In this paper, we present a variational study of the interactions between particlelike excitations in full microscopic detail. We construct stationary scattering states and calculate scattering phase shifts between particles with arbitrary individual momenta. Our method is applied to the spin-1 Heisenberg antiferromagnet, for which we calculate the full magnon-magnon S matrix, the two-particle contribution to the spectral function and higher order corrections to the magnetization curve.
\addcontentsline{toc}{section}{Variational method}
\tussentitel{Variational method.} Consider a one-dimensional spin system with local dimension $d$ in the thermodynamic limit, described by a local and translation invariant Hamiltonian $\ham=\sum_{n\in\ZZ}\hat{h}_{n,n+1}$, where we restrict to nearest neighbour interaction. The translation invariant ground state of this system can be accurately described by a uniform matrix product state \cite{Haegeman2011d,Fannes1992, *Vidal2007b, *Haegeman2014}
\begin{align*}
    \ket{\Psi[A]} = \sum_{\{s\}=1}^d \lv \left[ \prod_{m\in\ZZ} A^{s_m} \right] \rv \spst,
\end{align*}
where the $D\times d \times D$ tensor $A^s$ contains all variational parameters. 
\par Having found the ground state (by, \textit{e.g.}, simulating imaginary time evolution using the time-dependent variational principle \cite{Haegeman2011d}), the variational ansatz for an elementary excitation with definite momentum $\kappa$ is given by \cite{Haegeman2012a,Haegeman2013b}
\begin{multline} \label{OneParticle}
  \ket{\Phi_\kappa[B]} = \sum_{n=-\infty}^{+\infty} \e^{i\kappa n}\sum_{\{s\}=1}^d \lv \left[\prod_{m=-\infty}^{n-1} A^{s_m}\right] B^{s_n} \\
  \times \left[\prod_{m=n+1}^{+\infty} A^{s_m}\right] \rv \ket{\{s\}}
\end{multline}
and can be understood as a localized disturbance of an essentially unchanged ground state. Due to its matrix product representation, this localized disturbance can spread out over a distance determined by the bond dimension $D$. As the variational subspace of excited states defined by \eqref{OneParticle} is linear, finding the best approximation for the lowest lying excited states is achieved by solving an eigenvalue problem.
\par Because of the locality of the ansatz \eqref{OneParticle}, we can interpret the excitation as a particle and construct scattering states of two particlelike excitations. The variational ansatz for states with two elementary excitations with momenta $\kappa_1$ and $\kappa_2$ is taken to be ($\kappa=\kappa_1+\kappa_2$)
\begin{align} \label{ansatz}
  \ket{\Upsilon_{\kappa_1\kappa_2}} = \ket{\chi_{\kappa_1\kappa_2}(0)} + \sum_{\alpha\beta} \sum_{n=1}^{+\infty}c_{\kappa_1\kappa_2}^{\alpha\beta}(n) \ket{\chi_{\alpha\beta,\kappa}(n)}
\end{align}
with the states
\begin{widetext}
\begin{align}
   & \ket{\chi_{\kappa_1\kappa_2}(0)} = \sum_{n=-\infty}^{+\infty} \e^{i\kappa n} \sum_{\{s\}=1}^d \lv \left[ \prod_{m<n} A^{s_m} \right] C_{\kappa_1\kappa_2}^{s_n} \left[ \prod_{m>n} A^{s_m} \right] \rv \spst \label{local} \\
  & \ket{\chi_{\alpha\beta,\kappa}(n)} = \sum_{n_1=-\infty}^{+\infty} \e^{i\kappa n_1} \sum_{\{s\}=1}^d \lv \left[ \prod_{m<n_1} A^{s_m} \right] B_\alpha^{s_{n_1}} \left[ \prod_{n_1<m<n_1+n} A^{s_m} \right] B_\beta^{s_{n_1+n}} \left[ \prod_{m>n_1+n} A^{s_m} \right] \rv \spst. \label{nonLocal}
\end{align}
\end{widetext}
For the $B_{\alpha/\beta}$ in the states \eqref{nonLocal}, we use the two $B$-tensors that were found for the one-particle problem at momenta $\kappa_1$ and $\kappa_2$. This restriction is accurate when both particles are far away, but fails when the particles approach. The local term \eqref{local} should be able to correct for this, however, because of its ability to spread over some finite distance. For this reason, we keep all $D^2(d-1)$ variational parameters in $C^s_{\kappa_1\kappa_2}$.
\par Finding eigenstates within this (linear) variational subspace requires solving the generalized eigenvalue problem $H_{\text{eff}}\bar{c}=\omega N_{\text{eff}} \bar{c}$ with $\bar{c}=\left\{C_{\kappa_1\kappa_2}^s,c_{\kappa_1\kappa_2}^{\alpha\beta}(n) \right\}$ containing all variational parameters, $\omega$ being the total energy of the excitation (with the ground state energy $E_0$ subtracted) and an effective Hamiltonian and norm matrix given by
\begin{equation} \label{effective}
\begin{split}
  & H_{\text{eff}} = \bra{\chi(n)}\left( \hat{H}-E_0 \right) \ket{\chi(n')} \\
  & N_{\text{eff}} = \braket{\chi(n)|\chi(n')} .
\end{split}
\end{equation}
\par Finding solutions for this half-infinite eigenvalue problem with definite energy $\omega$ starts with an inspection of the asymptotic regime, \ie the regime where the two particles are considered to be infinitely far apart. For $n',n\rightarrow\infty$ the effective norm matrix is diagonal and the effective Hamiltonian matrix is reduced to repeating rows of block matrices that decay exponentially away from the diagonal. These blocks can be considered to be zero if they are, say, $N+1$ sites from the diagonal and for every set $(\kappa,\omega)$, we obtain a recurrence relation for the coefficients $c^{\alpha\beta}(n)$. This recurrence relation typically has two solutions with modulus one, which correspond to the incoming and outgoing plane waves with total momentum $\kappa$ and energy $\omega$, a large number of solutions within the unit circle, which correspond to decaying solutions as $n\rightarrow\infty$, and an equally large number of solutions with modulus larger than unity, which should be discarded as they are non-normalizable (\ie nonphysical) solutions to the eigenvalue problem.
\par We now construct solutions to the full eigenvalue problem that reduce to these asymptotic solutions for $n\rightarrow\infty$. They are obtained by writing the coefficients $c^{\alpha\beta}(n)$ as $c=Q x$, where $Q$ is a block diagonal of (i) a unit matrix of finite dimension, that leaves open the coefficients $c^{\alpha\beta}(n)$ for $n<N'$ and (ii) a matrix consisting of the asymptotic solutions, and $x$ is a new vector of coefficients. We thus assume that the asymptotic regime is reached when the two particles are a finite distance $N'$ apart; this implies that, when imposing $c=Q x$, the eigenvalue problem with energy $\omega$ is automatically fulfilled after $N'$ rows. Consequently, we can truncate the infinite set of equations and solve the system to find the finite-dimensional vector $x$. When this whole procedure is done consistently \cite{preparation}, an exact solution to this slightly different scattering problem is guaranteed to exist and, when the approximations are negligible, should give an approximate stationary scattering state with total momentum $\kappa$ and energy $\omega$. More specifically, the coefficients $c^{\alpha\beta}(n)$ for this state converge asymptotically to the form
\begin{multline} \label{scatteringPhase}
   c_{\kappa_1\kappa_2}^{\alpha\beta}(n) \overset{n\rightarrow\infty}{\rightarrow} u^\alpha(\kappa_1) u^\beta(\kappa_2) \e^{i\kappa_2n} \\ - \e^{i\phi} u^\alpha(\kappa_2)u^\beta(\kappa_1) \e^{i\kappa_1n},
\end{multline}
where $u^{\alpha/\beta}(\kappa)$ corresponds to the one-particle solution at momentum $\kappa$. This form allows for a direct calculation of the scattering phase $\phi$. \footnote{Thanks to the tensor structure of MPS, our method can be implemented with a computational complexity scaling as $\mathcal{O}(D^3)$ in the bond dimension.} 
\addcontentsline{toc}{section}{Application}
\tussentitel{Application.} The spin-1 Heisenberg antiferromagnet is defined by the Hamiltonian
\begin{align} \label{heisenberg}
  \ham = \sum_{n} \hat{S}^x_n\hat{S}^x_{n+1} + \hat{S}^y_n\hat{S}^y_{n+1} + \hat{S}^z_n\hat{S}^z_{n+1}.
\end{align}
Since Haldane's conjecture of the existence of a gap \cite{Haldane1983a}, the low-lying excitation spectrum  has been studied extensively \cite{Takahashi1989, *Takahashi1994, *White1993, *Sorensen1993, *Sorensen1994, *Sorensen1994a,  White2008a}. The spectrum has an isolated, threefold degenerate one-particle (magnon) branch centered around momentum $\pi$. At momentum $\kappa\approx0.22\pi$, this magnon triplet becomes unstable and a continuum of two-magnon scattering states emerges around momentum $0$.
\par Magnon interactions have been studied in Ref. \onlinecite{Lou2000}, where it was shown that the scattering of two magnons with individual momenta around $\pi$ can be parametrized by one parameter, the scattering length $a$. Indeed, for small momenta ($\kappa_{1,2}\rightarrow\pi$) the phase shift behaves as $\phi\left( \kappa_1,\kappa_2 \right) \approx - a \left( \kappa_1-\kappa_2 \right)$, hence the definition of $a$. Within the sector with total spin $S=2$, this quantity can be determined from the finite size correction to the energy of the lowest lying state within this sector. DMRG simulations have given an approximate value of $a_2\approx-2$ \cite{Lou2000} and (more recently) a more precise value of $a_2=-2.30(4)$ \cite{Ueda2011}. Through the identification of the Heisenberg chain with the non-linear sigma model (for which the full S matrix can be calculated exactly \cite{Zamolodchikov1979}), qualitative estimates of all three scattering lengths can be made (see Ref. \onlinecite{Lou2000}).
\par We now investigate the two-magnon scattering with our variational method. In Refs. \onlinecite{Haegeman2012a} and \onlinecite{Haegeman2013b} it was shown that the one-particle ansatz \eqref{OneParticle} is capable of describing the elementary magnon triplet with great precision. Now we can construct two-magnon states with every combination of individual momenta for which the magnon is stable (\ie $\left|\kappa_{1,2}\right|>0.22\pi$) and for every combination of individual spins. From the wave functions we can determine every phase shift and compute the full magnon-magnon S matrix. As the Hamiltonian \eqref{heisenberg} is $SU(2)$ invariant, we expect this S matrix to be diagonal in the coupled basis, with the matrix elements equal within each sector of total spin.
\begin{figure}
	\includegraphics[width=\columnwidth]{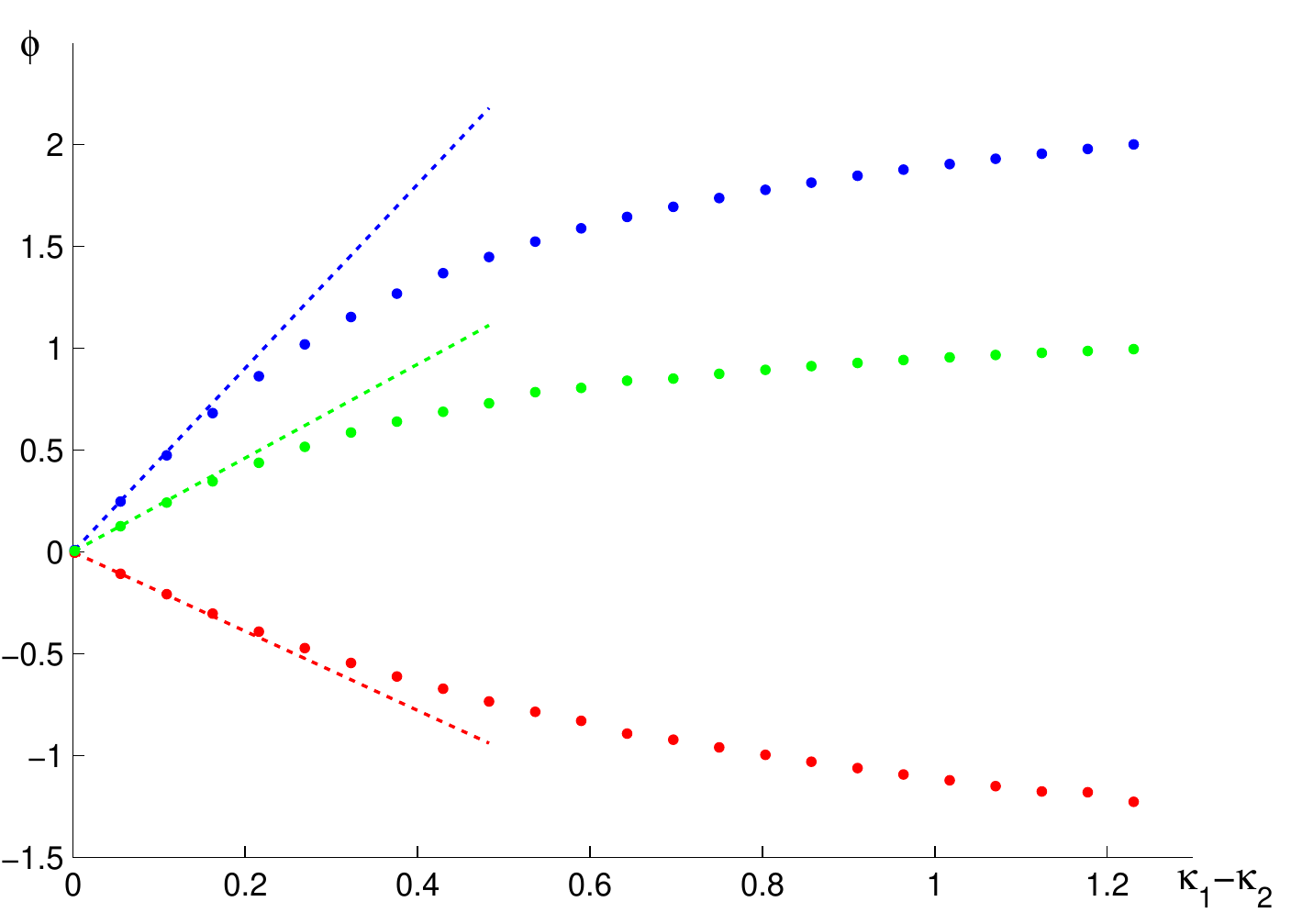}
    \caption{\label{phase}The angle of the S-matrix elements for the sectors of total spin $S=2$ (green), $S=1$ (blue) and $S=0$ (red) at total momentum $\kappa=0$ and different relative momenta $\kappa_\text{rel}=\kappa_1-\kappa_2$. The linear regime for small relative momenta is clearly visible, as well as the deviations from that regime at higher relative momenta. Calculations were done with bond dimension $D=64$.}
\end{figure}
\par In Fig. \ref{phase} we have plotted the scattering phases within each spin sector for different relative momenta (our method reproduces the block structure of the S matrix, so all information is contained in these three phases). We can clearly observe a linear regime where the relative momentum is small, with the slope giving us a direct measure of the scattering length in the different sectors. We find the following values for the scattering lengths \footnote{We refer to the supplemental material for more details on the precision of these values.}
\begin{align*}
  a_0 = 1.945 \qquad a_1 = -4.515 \qquad a_2 = -2.306.
\end{align*}
The signs of these scattering lengths are in agreement with the predictions of the non-linear sigma model. In the $S=2$ sector we have excellent agreement with Ref. \onlinecite{Ueda2011}, while for the other sectors we have found no previous quantitative estimates.
\par When we go to larger relative momenta, the curve loses its linearity. In this regime, the low-energy description of the scattering process in terms of the scattering length is no longer valid and the S matrix can only be determined by solving the full microscopic scattering problem. Since the effective Hamiltonian \eqref{effective} of the scattering problem indeed captures the microscopic details of the magnon-magnon interaction, our method is able to study scattering in this nontrivial regime also.
\par
\begin{figure}
	\includegraphics[width=\columnwidth]{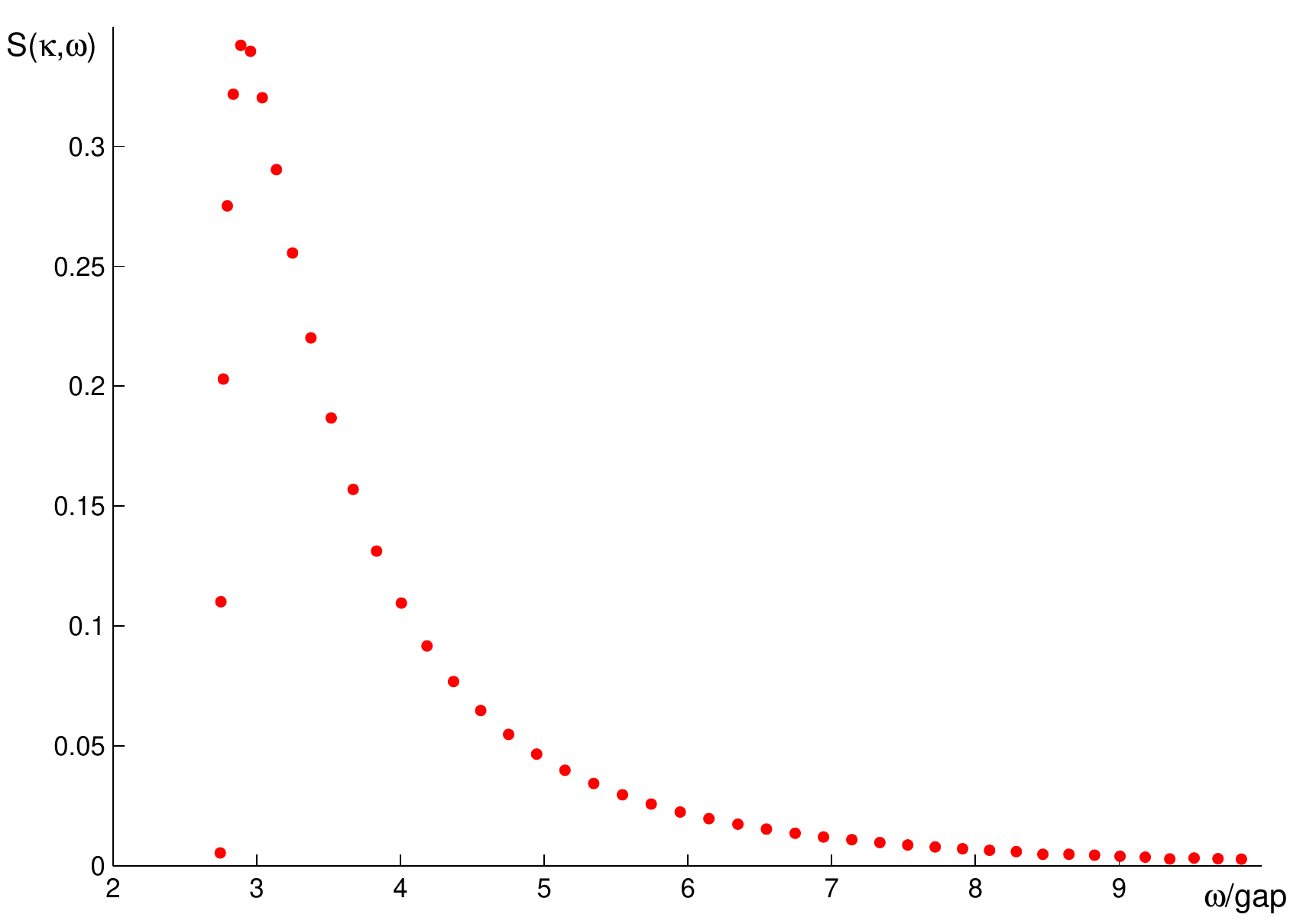}
    \caption{\label{spectral}The two-particle contribution for the spectral function $S(\kappa,\omega)$ at momentum $\kappa=\frac{\pi}{10}$. Calculations were done with bond dimension $D=48$.}
\end{figure}

 Next we turn to the spectral function, defined as
\begin{align*}
S(\kappa,\omega) = \sum_{n=-\infty}^{+\infty} \e^{-i\kappa n} \int_{-\infty}^{+\infty} \d t \, \e^{i\omega t} \bra{\Psi_0} S^y_n(t) S^y_0(0) \ket{\Psi_0}.
\end{align*}
Since we have constructed the wave function of all two-particle states explicitly, we can calculate their spectral weights and, consequently, the two-particle contribution to $S(\kappa,\omega)$. This contribution is expected to be dominant around momentum zero, as the two-particle states are the lowest lying excited states in that region \cite{Affleck1992}. In Fig. \ref{spectral} we have plotted the spectral function at momentum $\kappa=\frac{\pi}{10}$. Comparing our results with Ref. \cite{White2008a}, where the spectral function was calculated using DMRG techniques for real-time evolution and linear prediction, shows that we are able to capture the two-particle states perfectly.
\par We can get an idea of how well the full spectral function is reproduced by looking at its zeroth and first frequency moment, \ie $s_0(\kappa)=\int\frac{\d\omega}{2\pi} S(\kappa,\omega)$ and $s_1(\kappa)=\int\frac{\d\omega}{2\pi} \omega S(\kappa,\omega)$. As the former is equal to the static structure factor and the latter can be written as the expectation value of a simple double commutator \cite{Hohenberg1974}, both can be easily calculated with the MPS ground state. It appears that the two-particle contribution in Fig. \ref{spectral} approaches the exact values up to 98.7\% and 96.4\%, showing indeed that the two-particle sector carries the dominant contribution of the spectral function at this momentum. Note that, as our method relies on the explicit wave function of the excitations directly in the thermodynamic limit, our results do not suffer from finite size effects nor statistical errors.
\par As another application, we use our variational results for the magnon dispersion and the magnon-magnon S matrix to study the magnetization of the Heisenberg chain when applying a critical magnetic field. In previous publications, the finite density of magnons has been described as a gas of interacting bosons \cite{Affleck1991} with quadratic dispersion, for which the scattering length gives a first-order correction to the hard-core boson description \cite{Lou2000}. The magnetized chain has been characterized as a Luttinger liquid (LL) \cite{Giamarchi2004} with a LL parameter that varies with the magnetization \cite{Fath2003, Konik2002, Affleck2005}.
\par
\begin{figure}
	\includegraphics[width=\columnwidth]{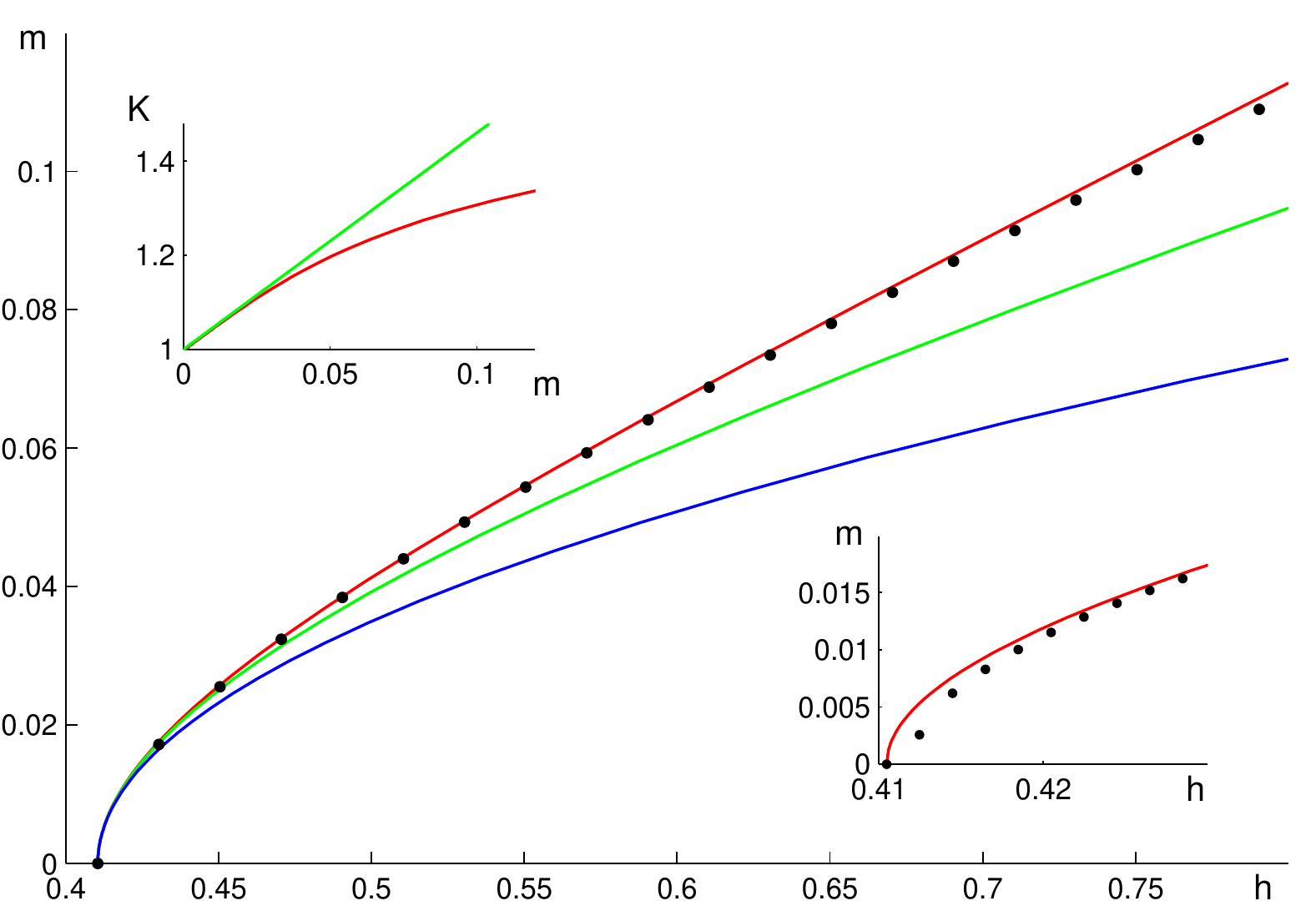}
    \caption{\label{magnetization}The magnetization $m$ versus applied magnetic field $h$ for the spin-1 Heisenberg chain. Our results (red, bond dimension $D=64$) are compared to the hard-core boson square-root dependence (blue) \cite{Affleck1991} and first order corrections by the scattering length $a_2$ (green) \cite{Lou2000}. The direct MPS calculations (black dots) were done at the same bond dimension of $D=64$. The bottom-right inset provides a close-up of the phase transition. The top-left inset provides our result for the LL parameter $K$ in function of the magnetization $m$ (red), compared to the linear relation based on the scattering length (green) \cite{Affleck2005}.}
\end{figure}

 As the present method provides complete information on two-magnon interactions, we can use this to approximately describe the finite density of magnons. Indeed, we can neglect three-particle interactions and write down the Bethe ansatz wave function \cite{Bethe1931, *Korepin1997} (with the variationally determined phase shifts) as an approximation of the true wave function of the magnon gas. Solving the corresponding Bethe equations (with our variationally determined dispersion relation) numerically \footnote{We refer to the supplemental material for more details on this procedure.}, the magnetization curve as well as the LL parameter can be obtained (see Fig. \ref{magnetization}). We expect this to be a good approximation at low magnon densities, where three-magnon interactions are negligible. A comparison with direct MPS calculations shows that our description is indeed very accurate in a broad regime and does not share the difficulties of traditional DMRG/MPS methods for capturing the onset of criticality.
\addcontentsline{toc}{section}{Conclusions and outlook}
\tussentitel{Conclusions and outlook.} Starting from a successful particlelike ansatz for elementary excitations, we introduced a variational method for constructing two-particle states and determining their scattering phase shifts and spectral weights. This information was then used to determine the critical properties of a finite density of these excitations. We believe that our methods open up new routes towards a better understanding of the low-lying dynamics of (quasi-) one-dimensional quantum spin systems.
\par Indeed, our methods can be straightforwardly applied to more interesting systems such as \eg spin ladders and dimerized chains beyond the strong-coupling limit \footnote{In the supplemental material we provide some preliminary results on the application to a spin ladder.}. Our formalism can be extended to topologically non-trivial excitations, so we can study \eg spinon interactions in half-integer spin chains, and bound states, which correspond to solutions of the scattering problem without any non-decaying asymptotic solutions. We might also study systems at finite temperature, using semiclassical approximations \cite{Damle1998}, the thermodynamic Bethe ansatz and/or form-factor expansions \cite{Essler2008}.
\par Most importantly, as we have shown to give an accurate microscopic description of the interactions of elementary excitations, we are able to build an effective theory of interacting particlelike excitations for capturing the low-energy physics of generic spin chains. By gradually averaging out the microscopic details of the interactions, we can systematically make the connection to previous effective field theories based on phenomenological considerations and symmetries, globally determined parameters and/or strong- or weak-coupling limits.
\tussentitel{Acknowledgments.} We acknowledge support from an Odysseus grant from the Research Foundation Flanders, the EU grants QUERG, SIQS, and by the Austrian FWF SFB grants FoQus and ViCoM. L.V. is supported by a Doctoral Scholarship from the Research Foundation Flanders. T.J.O. is supported by the EU grant QFTCMPS and the the cluster of excellence EXC 201 Quantum Engineering and Space-Time Research.

\newpage
\appendix
\onecolumngrid
\begin{center} {\large \textbf{Supplemental Material}} \end{center}
\vspace{1.5cm}

\twocolumngrid

\section{Scattering length}

In the main body of the article, we provided our variational estimates for the magnon-magnon scattering lengths in all three total spin sectors for the spin-1 Heisenberg antiferromagnet. In this supplemental material we provide more detailed data on the basis of which we were able to estimate the precision for the given values.
\par First of all, we comment on the two parameters $N$ and $N'$ that were introduced to solve the scattering problem. As all matrix elements in the effective Hamiltonian and norm matrix decay exponentially when going away from the diagonal, \ie $\bra{\chi(n')}\hat{H}\ket{\chi(n)}\propto \e^{-|n'-n|/\xi}$ with $\xi$ the correlation length of the ground state. For the spin-1 Heisenberg chain $N\approx70$ gives quantitatively correct results. The right value of the parameter $N'$ is somewhat harder to predict, but from our simulations it is clear that from $N'\approx100$ the results are quantitatively correct. As these values are easily attained, no systematic errors are produced from having to work with finite $N$ and $N'$.
\par
\begin{figure}[b]
	\includegraphics[width=1\columnwidth]{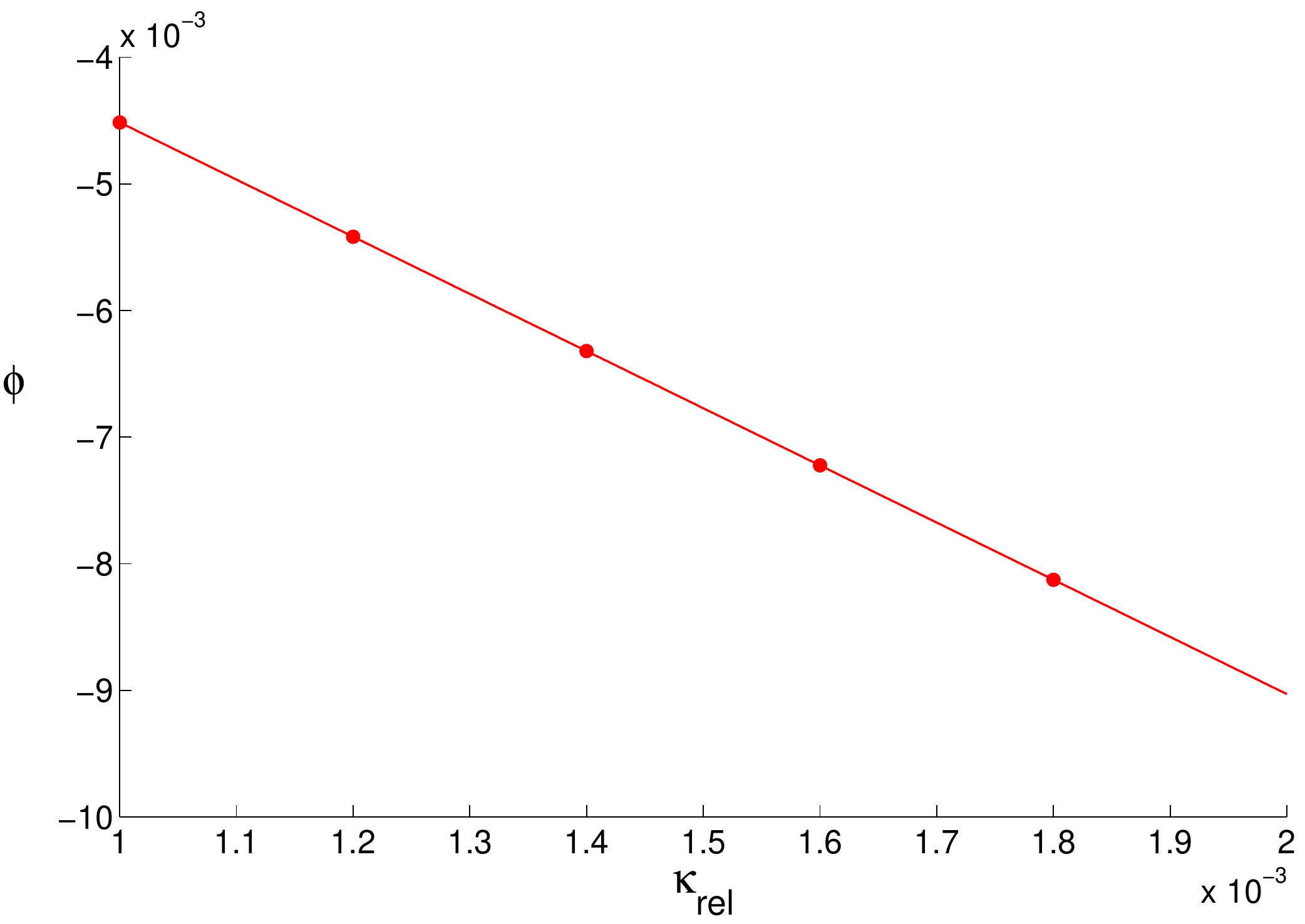}
    \caption{\label{slope}The scattering phase for small relative momenta in the sector with total spin $S=1$ with bond dimension $D=220$. The slope of the linear fit is given in the above table, the intercept has absolute value $7.3\times10^{-8}$ and the norm of the residuals is $8.63\times10^{-9}$.}
\end{figure}

 A second possible source of error is the linear fit for the scattering phase. The scattering length is determined as the slope of the scattering phase $\phi$ in function of the relative momentum $\kappa_\text{rel}=\kappa_1-\kappa_2$. In Fig. \ref{slope} we plot $\phi$ for very small relative momenta, showing that the linearity is reproduced up to very high precision. We can safely say that no significant error results from doing the linear fit.
\par In all matrix product state calculations the parameter that controls the accuracy of the variational approximation is the bond dimension $D$. To get an estimate of the variational error, we should compare results for the scattering lengths for different $D$ (see table \ref{scatLength}). Since there is no variational principle for the scattering length in our framework, we do not expect monotonic behaviour.
\begin{table}[b]
\centering
\begin{tabular}{|c|c|c|c|}
\hline
 & $a_0$ & $a_1$ & $a_2$ \\
 \hline
 $D=120$ & 1.94475 & -4.51330 & -2.35951 \\
 \hline
 $D=142$ & 1.94777 & -4.51535 & -2.30559 \\
 \hline
 $D=162$ & 1.94493 & -4.51561 & -2.30491 \\
 \hline
 $D=192$ & 1.94454 & -4.51527 & -2.30586 \\
 \hline
 $D=208$ & 1.94470 & -4.50912 & -2.30587 \\
 \hline
 $D=220$ & 1.94492 & -4.51537 & -2.30598 \\
 \hline \hline
 & 1.945 & -4.515 & -2.306 \\
 \hline
\end{tabular}
\caption{\label{scatLength}Convergence of the scattering length for different values of the MPS bond dimension $D$.}
\end{table}
\par We can see that these values are converged up to the reported precision in the last line, up to a strongly deviating value in the $S=1$ sector for $D=208$. We have no adequate explanation for this deviation.

\section{Magnetization curve}

It is well known that, when applying a magnetic field $h$ to the Heisenberg chain, the ground state is unaffected until the field reaches a critical value $h_c$ equal to the mass gap $\Delta$. Beyond this critical field a density of magnons is formed, for which the magnetic field serves as a chemical potential $\mu=h-\Delta$ (magnon interactions keep the density finite). 
\par The magnetization curve in the main body of the article was obtained by writing down the Bethe ansatz wave function as an approximate description of the finite density of magnons, and solving the corresponding Bethe equations numerically \cite{Korepin1997}. In first quantization (system with $N$ particles), this wave function can be written as
\begin{equation*}
\Psi(x_1,x_2,\dots,x_N) = \sum_{\perm} A(\perm) \e^{i(\lambda_{\perm 1}x_1 + \lambda_{\perm 2}x_2 + \dots+\lambda_{\perm N}x_N)} 
\end{equation*}
where $A(\perm)/A(\perm') = S(\lambda_i,\lambda_j)$ if the permutations $\perm$ and $\perm'$ differ by the interchange of the momenta $\lambda_i$ and $\lambda_j$; $S(\lambda_i,\lambda_j)$ is the S matrix for the scattering of two particles with momenta $\lambda_i$ and $\lambda_j$, which we can calculate with our variational method.
\par Imposing periodic boundary conditions on this wave function leads to the Bethe equations, which, upon going to the thermodynamic limit, is transformed into a linear integral equation for the density of particles in momentum space
\begin{equation*}
\rho(\lambda) -\frac{1}{2\pi} \int_{-q}^{q} K(\lambda,\lambda') \rho(\lambda') \d\lambda' = \frac{1}{2\pi},
\end{equation*}
where the kernel of the integral equation is given by the derivative of the scattering phase, \ie $K(\lambda,\lambda')=\partial_\lambda\phi(\lambda,\lambda')$, and $q$ is the Fermi momentum. \footnote{In textbook derivations of the Bethe equations the scattering phase is often taken to be only dependent on the difference of the two individual momenta, because of Galilean invariance. For the magnetized chain this will only be approximately the case. For the sake of simplicity we have assumed this to be the case, however, and it remains a subject for further study to track the degree and implications of the breaking of Galilean invariance.}
\par In the grand-canonical ensemble (with chemical potential $\mu$), a second integral equation can be obtained for the dressed energy function $\epsilon(\lambda)$
\begin{equation*}
\epsilon(\lambda) -\frac{1}{2\pi} \int_{-q}^{q} K(\lambda,\lambda') \epsilon(\lambda') \d\lambda' = \epsilon_0(\lambda) - \mu,
\end{equation*}
where $\epsilon_0(\lambda)$ is the bare (kinetic) energy of a magnon with momentum $\lambda$, which we also calculate variationally. The value of the Fermi momentum is fixed by demanding that $\epsilon(q)=0$ and can be obtained numerically. The density of magnons (\ie the magnetization) is then calculated with
\begin{equation*}
D = \int_{-q}^{q} \rho(\lambda)\d\lambda.
\end{equation*}
\par Luttinger liquid parameters for the magnetized spin chain can be obtained by calculating the sound velocity and the compressibility based on this approximate Bethe ansatz description. We can determine the sound velocity of the system 
\begin{equation*}
v_S = \left. \frac{\partial e(\lambda)}{\partial p(\lambda)} \right|_{p(\lambda)=q}
\end{equation*}
by calculating the momentum and energy of elementary excitations close to the Fermi level. A particle excitation with Bethe momentum $\lambda>q$ has physical momentum
\begin{equation*}
p(\lambda) = \lambda + \int_{-q}^{q} \phi(\lambda,\mu) \rho(\mu) \d\mu
\end{equation*}
and energy
\begin{equation*}
\e(\lambda) = \epsilon_0(\lambda) + \frac{1}{2\pi} \int_{-q}^{q} K(\lambda,\mu) \epsilon(\mu) \d\mu.
\end{equation*}
The compressibility can be calculated easily as
\begin{equation*}
\kappa = \frac{1}{D^2} \frac{\partial D}{\partial \mu}.
\end{equation*}
The sound velocity and the compressibility can be directly linked to the two Luttinger liquid parameters \cite{Giamarchi2004}.

\section{Spin ladders}

In this section we present some preliminary results for the application of our methods to spin ladders. We will look at the two-leg antiferromagnetic spin-1/2 ladder, given by the Hamiltonian
\begin{equation*}
H = J_\parallel \sum_{n,l}  \vec{S}_{n,l} \cdot \vec{S}_{n+1,l} + J_\perp \sum_n  \vec{S}_{n,1} \cdot \vec{S}_{n,2}
\end{equation*}
where $l=1,2$ denote the two legs of the ladder and the $\vec{S}$ are spin-1/2 operators. The model is determined by the ratio $\gamma=J_\perp/J_\parallel$ . The lowest-lying elementary excitation is again a triplet with mimimum around $\pi$, so we can study the scattering of these particles. Depending on the parameter $\gamma$ bound states may form.
\par
\begin{figure}
	\includegraphics[width=1\columnwidth]{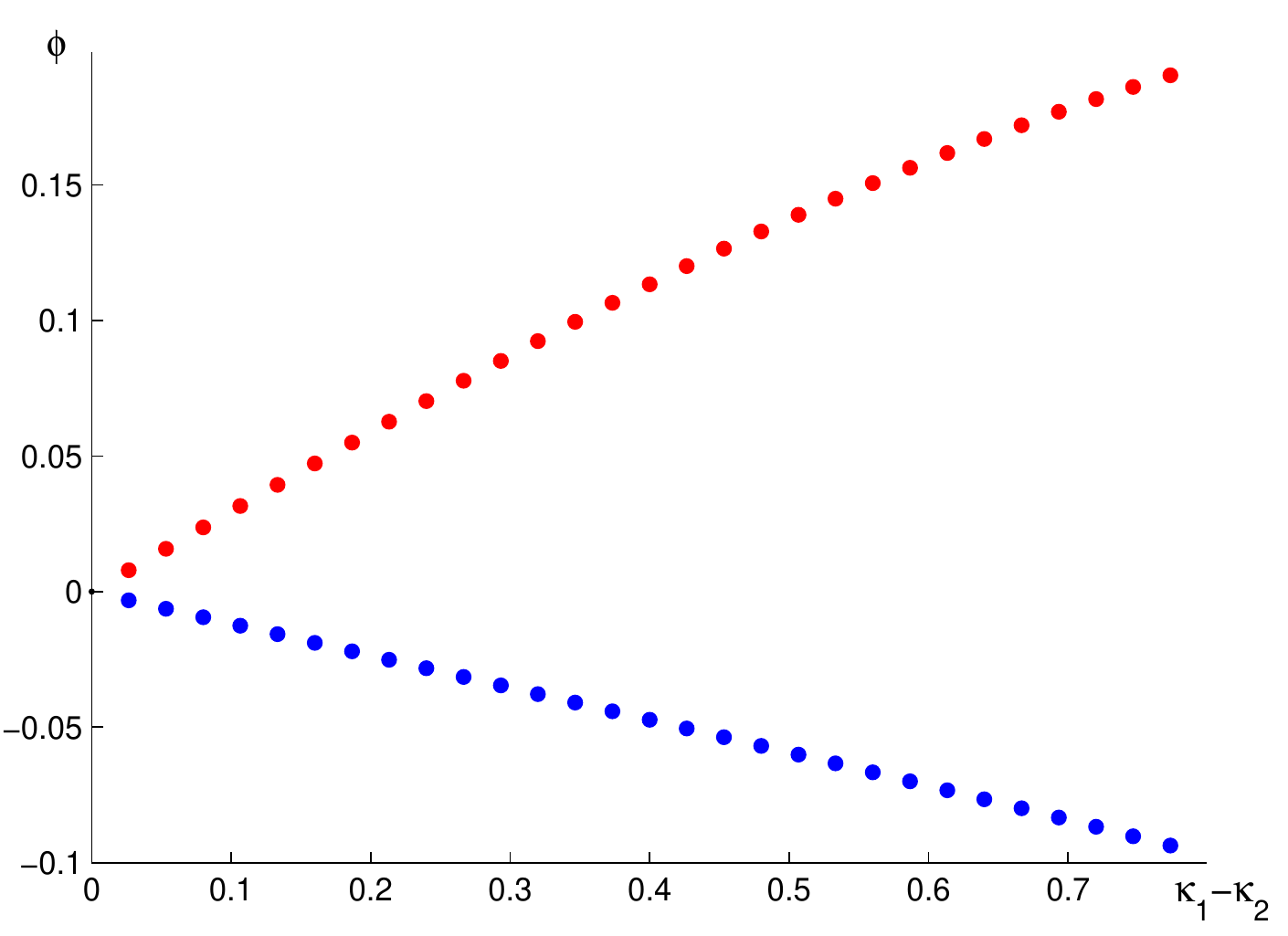}
    \caption{\label{ladder1} The scattering phase $\phi$ of the spin-1/2 ladder with $\gamma=1.5$ (red) and $\gamma=3.5$ (blue) at total momentum $\kappa=0$ and for different values of the relative momentum $\kappa_1-\kappa_2$.}
\end{figure}

\begin{figure*}[t!]
\centering
\subfigure{\includegraphics[width=0.8\columnwidth]{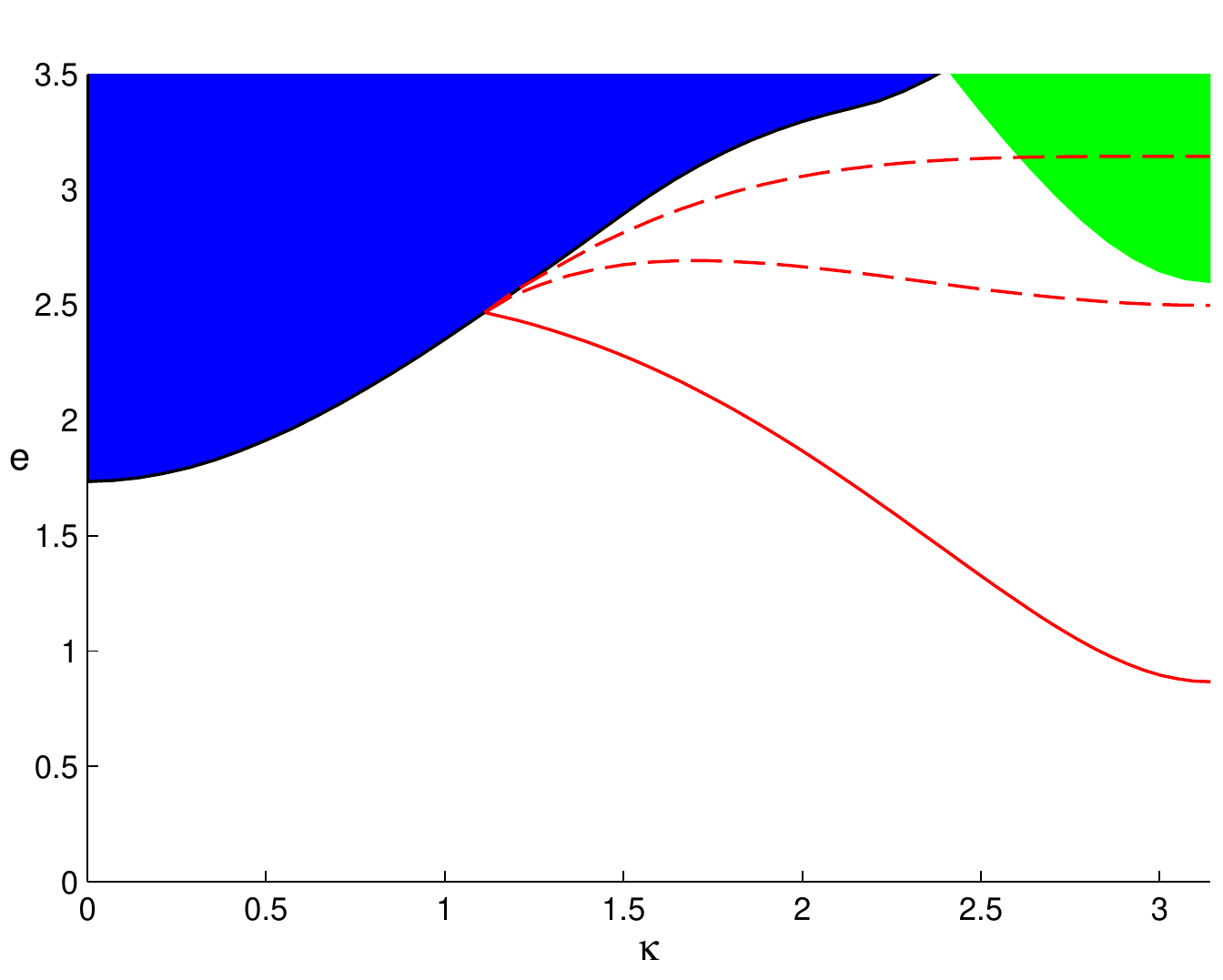} \label{fig:dispersion1}} \qquad
\subfigure{\includegraphics[width=0.8\columnwidth]{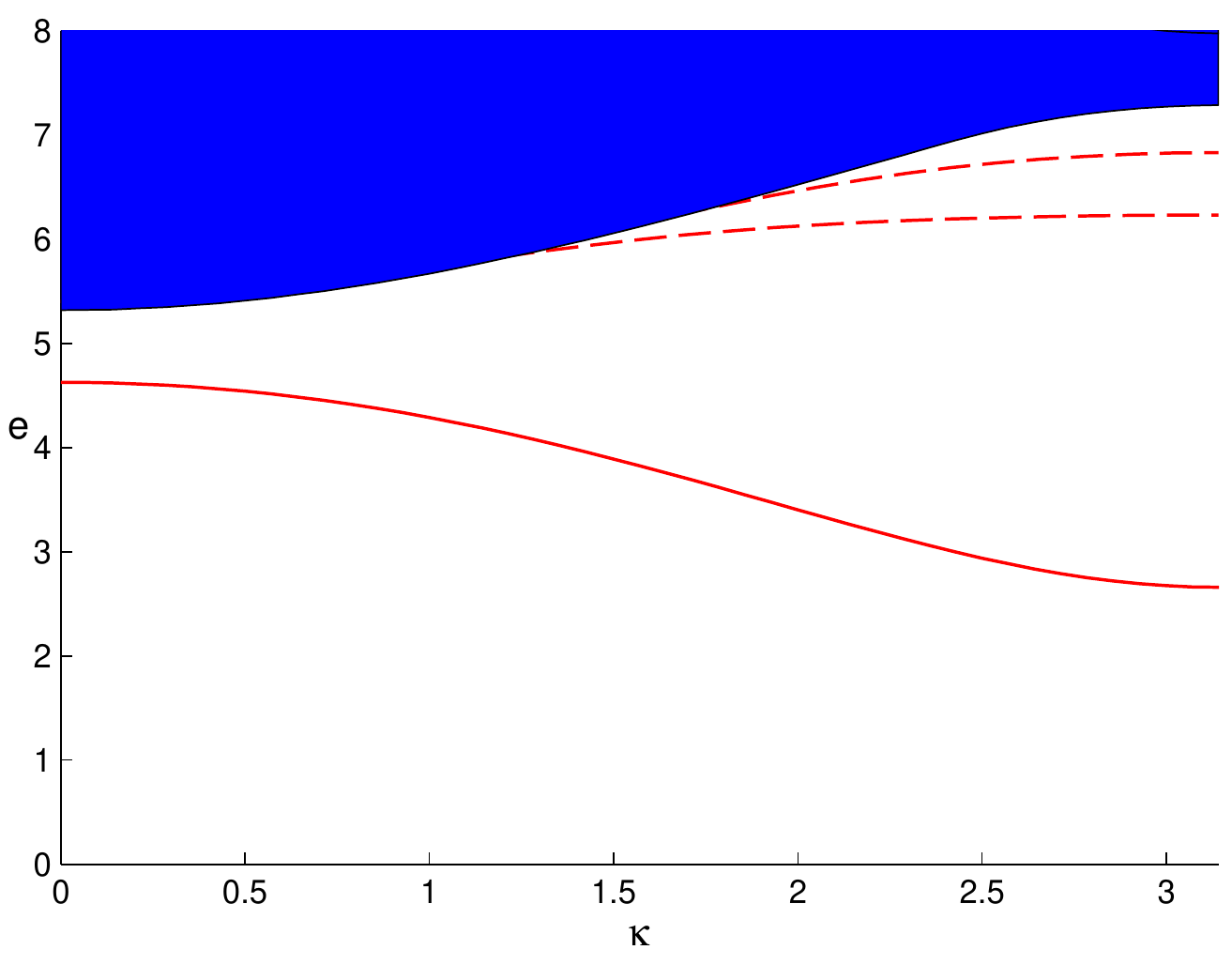} \label{fig:dispersion2}}
\caption{\label{dispersion}Elementary excitation spectrum of the spin-1/2 ladder for $\gamma=1.5$ (left) and $\gamma=3.5$ (right) as calculated with our variational methods: the elementary spin-1 triplet (full red curve), spin-0 and spin-1 bound states (striped red curves), two-particle continuum (blue) and three-particle continuum (green). On the left the spin-1 triplet becomes unstable into two-particle decay for momenta smaller than $\kappa\approx1.112$, while on the right it is stable for all momenta. On the right the three-particle continuum is further up and hence not shown.}
\end{figure*}

\begin{figure}
	\includegraphics[width=1\columnwidth]{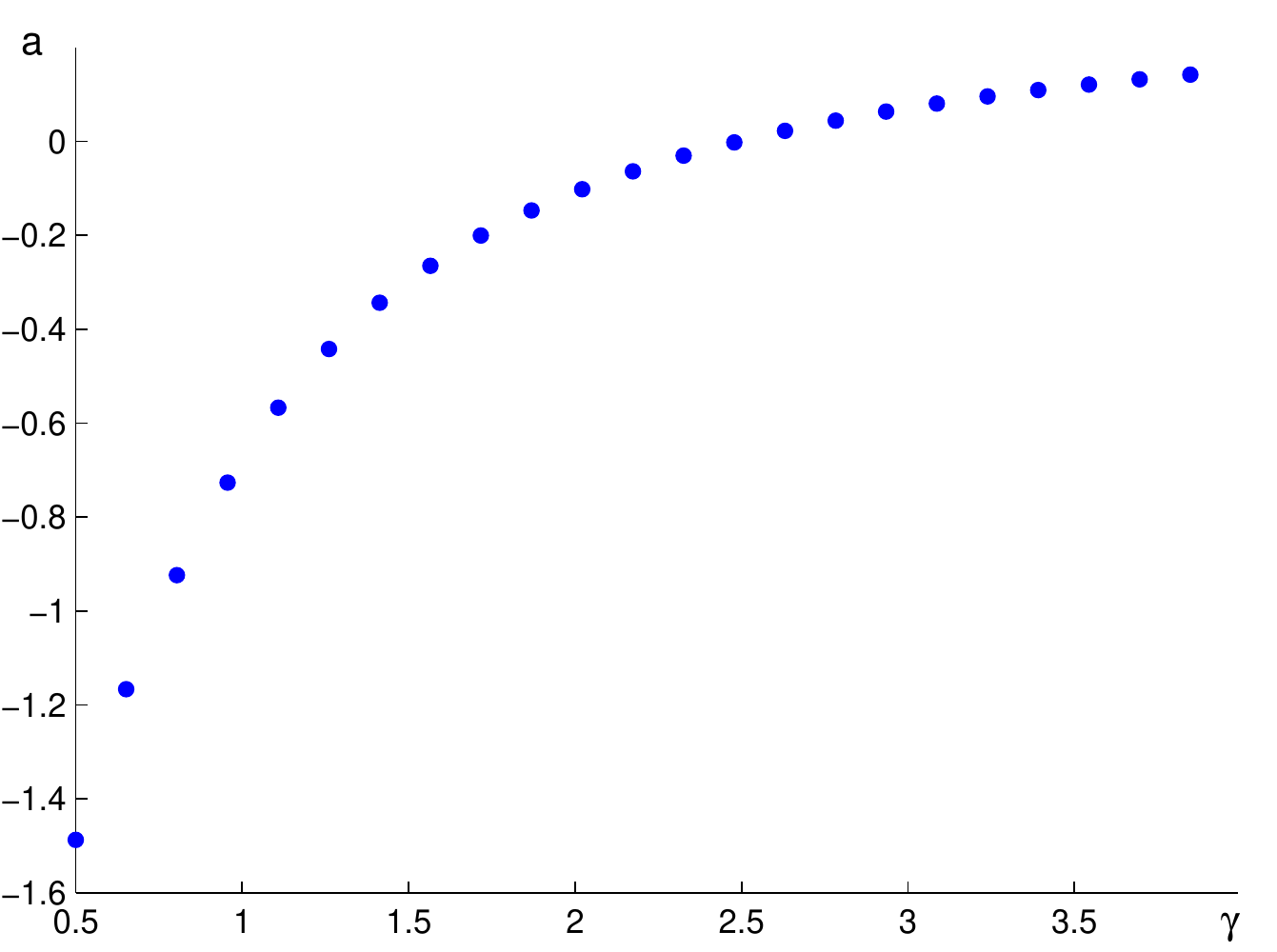}
    \caption{\label{ladder2}The scattering length of the spin-1/2 ladder for different values of the parameter ratio $\gamma$.}
\end{figure}

 In Fig. \ref{dispersion} the elementary excitation spectrum is shown for two specific values for $\gamma$ and in Fig. \ref{ladder1} we have plotted the phase shifts for two excitations within the $S=2$ sector.
As the scattering length determines leading-order corrections to some interesting properties, we calculated this quantity for different values of the parameter ratio $\gamma$. The results in Fig. \ref{ladder2} show that the scattering length switches sign as the ratio increases, which has far-reaching implications on, \eg, the Luttinger parameter for the ladder in a magnetic field.
\par In future work we intend to completely characterize the elementary excitation spectrum of the spin ladder, including bound states and scattering states, and calculate spectral weights. The S matrix will provide important information on the properties of the ladder when an external magnetic field is applied. \cite{preparation}


\begin{thebibliography}{59}%
\makeatletter
\providecommand \@ifxundefined [1]{%
 \@ifx{#1\undefined}
}%
\providecommand \@ifnum [1]{%
 \ifnum #1\expandafter \@firstoftwo
 \else \expandafter \@secondoftwo
 \fi
}%
\providecommand \@ifx [1]{%
 \ifx #1\expandafter \@firstoftwo
 \else \expandafter \@secondoftwo
 \fi
}%
\providecommand \natexlab [1]{#1}%
\providecommand \enquote  [1]{``#1''}%
\providecommand \bibnamefont  [1]{#1}%
\providecommand \bibfnamefont [1]{#1}%
\providecommand \citenamefont [1]{#1}%
\providecommand \href@noop [0]{\@secondoftwo}%
\providecommand \href [0]{\begingroup \@sanitize@url \@href}%
\providecommand \@href[1]{\@@startlink{#1}\@@href}%
\providecommand \@@href[1]{\endgroup#1\@@endlink}%
\providecommand \@sanitize@url [0]{\catcode `\\12\catcode `\$12\catcode
  `\&12\catcode `\#12\catcode `\^12\catcode `\_12\catcode `\%12\relax}%
\providecommand \@@startlink[1]{}%
\providecommand \@@endlink[0]{}%
\providecommand \url  [0]{\begingroup\@sanitize@url \@url }%
\providecommand \@url [1]{\endgroup\@href {#1}{\urlprefix }}%
\providecommand \urlprefix  [0]{URL }%
\providecommand \Eprint [0]{\href }%
\providecommand \doibase [0]{http://dx.doi.org/}%
\providecommand \selectlanguage [0]{\@gobble}%
\providecommand \bibinfo  [0]{\@secondoftwo}%
\providecommand \bibfield  [0]{\@secondoftwo}%
\providecommand \translation [1]{[#1]}%
\providecommand \BibitemOpen [0]{}%
\providecommand \bibitemStop [0]{}%
\providecommand \bibitemNoStop [0]{.\EOS\space}%
\providecommand \EOS [0]{\spacefactor3000\relax}%
\providecommand \BibitemShut  [1]{\csname bibitem#1\endcsname}%
\let\auto@bib@innerbib\@empty
\bibitem [{\citenamefont {Hastings}(2007)}]{Hastings2007}%
  \BibitemOpen
  \bibfield  {author} {\bibinfo {author} {\bibfnamefont {M.~B.}\ \bibnamefont
  {Hastings}},\ }\href {\doibase 10.1088/1742-5468/2007/08/P08024} {\bibfield
  {journal} {\bibinfo  {journal} {Journal of Statistical Mechanics: Theory and
  Experiment}\ }\textbf {\bibinfo {volume} {2007}},\ \bibinfo {pages} {P08024}
  (\bibinfo {year} {2007})}\BibitemShut {NoStop}%
\bibitem [{\citenamefont {Masanes}(2009)}]{Masanes2009}%
  \BibitemOpen
  \bibfield  {author} {\bibinfo {author} {\bibfnamefont {L.}~\bibnamefont
  {Masanes}},\ }\href {\doibase 10.1103/PhysRevA.80.052104} {\bibfield
  {journal} {\bibinfo  {journal} {Physical Review A}\ }\textbf {\bibinfo
  {volume} {80}},\ \bibinfo {pages} {052104} (\bibinfo {year}
  {2009})}\BibitemShut {NoStop}%
\bibitem [{\citenamefont {Schollw\"{o}ck}(2011)}]{Schollwock2011a}%
  \BibitemOpen
  \bibfield  {author} {\bibinfo {author} {\bibfnamefont {U.}~\bibnamefont
  {Schollw\"{o}ck}},\ }\href {\doibase 10.1016/j.aop.2010.09.012} {\bibfield
  {journal} {\bibinfo  {journal} {Annals of Physics}\ }\textbf {\bibinfo
  {volume} {326}},\ \bibinfo {pages} {96} (\bibinfo {year} {2011})}\BibitemShut
  {NoStop}%
\bibitem [{\citenamefont {White}(1992)}]{White1992}%
  \BibitemOpen
  \bibfield  {author} {\bibinfo {author} {\bibfnamefont {S.~R.}\ \bibnamefont
  {White}},\ }\href {http://link.aps.org/doi/10.1103/PhysRevLett.69.2863}
  {\bibfield  {journal} {\bibinfo  {journal} {Physical Review Letters}\
  }\textbf {\bibinfo {volume} {69}},\ \bibinfo {pages} {2863} (\bibinfo {year}
  {1992})}\BibitemShut {NoStop}%
\bibitem [{\citenamefont {Hallberg}(1995)}]{Hallberg1995}%
  \BibitemOpen
  \bibfield  {author} {\bibinfo {author} {\bibfnamefont {K.}~\bibnamefont
  {Hallberg}},\ }\href {\doibase 10.1103/PhysRevB.52.R9827} {\bibfield
  {journal} {\bibinfo  {journal} {Physical Review B}\ }\textbf {\bibinfo
  {volume} {52}},\ \bibinfo {pages} {R9827} (\bibinfo {year}
  {1995})}\BibitemShut {NoStop}%
\bibitem [{\citenamefont {K\"{u}hner}\ and\ \citenamefont
  {White}(1999)}]{Kuhner1999}%
  \BibitemOpen
  \bibfield  {author} {\bibinfo {author} {\bibfnamefont {T.}~\bibnamefont
  {K\"{u}hner}}\ and\ \bibinfo {author} {\bibfnamefont {S.}~\bibnamefont
  {White}},\ }\href {\doibase 10.1103/PhysRevB.60.335} {\bibfield  {journal}
  {\bibinfo  {journal} {Physical Review B}\ }\textbf {\bibinfo {volume} {60}},\
  \bibinfo {pages} {335} (\bibinfo {year} {1999})}\BibitemShut {NoStop}%
\bibitem [{\citenamefont {Jeckelmann}(2002)}]{Jeckelmann2002}%
  \BibitemOpen
  \bibfield  {author} {\bibinfo {author} {\bibfnamefont {E.}~\bibnamefont
  {Jeckelmann}},\ }\href {\doibase 10.1103/PhysRevB.66.045114} {\bibfield
  {journal} {\bibinfo  {journal} {Physical Review B}\ }\textbf {\bibinfo
  {volume} {66}},\ \bibinfo {pages} {045114} (\bibinfo {year}
  {2002})}\BibitemShut {NoStop}%
\bibitem [{\citenamefont {Verstraete}\ \emph {et~al.}(2004)\citenamefont
  {Verstraete}, \citenamefont {Garc\'{\i}a-Ripoll},\ and\ \citenamefont
  {Cirac}}]{Verstraete2004c}%
  \BibitemOpen
  \bibfield  {author} {\bibinfo {author} {\bibfnamefont {F.}~\bibnamefont
  {Verstraete}}, \bibinfo {author} {\bibfnamefont {J.~J.}\ \bibnamefont
  {Garc\'{\i}a-Ripoll}}, \ and\ \bibinfo {author} {\bibfnamefont {J.~I.}\
  \bibnamefont {Cirac}},\ }\href {\doibase 10.1103/PhysRevLett.93.207204}
  {\bibfield  {journal} {\bibinfo  {journal} {Physical Review Letters}\
  }\textbf {\bibinfo {volume} {93}},\ \bibinfo {pages} {207204} (\bibinfo
  {year} {2004})}\BibitemShut {NoStop}%
\bibitem [{\citenamefont {Vidal}(2004)}]{Vidal2004a}%
  \BibitemOpen
  \bibfield  {author} {\bibinfo {author} {\bibfnamefont {G.}~\bibnamefont
  {Vidal}},\ }\href {\doibase 10.1103/PhysRevLett.93.040502} {\bibfield
  {journal} {\bibinfo  {journal} {Physical Review Letters}\ }\textbf {\bibinfo
  {volume} {93}},\ \bibinfo {pages} {4} (\bibinfo {year} {2004})}\BibitemShut
  {NoStop}%
\bibitem [{\citenamefont {White}\ and\ \citenamefont
  {Feiguin}(2004)}]{White2004}%
  \BibitemOpen
  \bibfield  {author} {\bibinfo {author} {\bibfnamefont {S.~R.}\ \bibnamefont
  {White}}\ and\ \bibinfo {author} {\bibfnamefont {A.~E.}\ \bibnamefont
  {Feiguin}},\ }\href {\doibase 10.1103/PhysRevLett.93.076401} {\bibfield
  {journal} {\bibinfo  {journal} {Physical Review Letters}\ }\textbf {\bibinfo
  {volume} {93}},\ \bibinfo {pages} {076401} (\bibinfo {year}
  {2004})}\BibitemShut {NoStop}%
\bibitem [{\citenamefont {Weichselbaum}\ \emph {et~al.}(2009)\citenamefont
  {Weichselbaum}, \citenamefont {Verstraete}, \citenamefont {Schollw\"{o}ck},
  \citenamefont {Cirac},\ and\ \citenamefont {von Delft}}]{Weichselbaum2009a}%
  \BibitemOpen
  \bibfield  {author} {\bibinfo {author} {\bibfnamefont {A.}~\bibnamefont
  {Weichselbaum}}, \bibinfo {author} {\bibfnamefont {F.}~\bibnamefont
  {Verstraete}}, \bibinfo {author} {\bibfnamefont {U.}~\bibnamefont
  {Schollw\"{o}ck}}, \bibinfo {author} {\bibfnamefont {J.~I.}\ \bibnamefont
  {Cirac}}, \ and\ \bibinfo {author} {\bibfnamefont {J.}~\bibnamefont {von
  Delft}},\ }\href {\doibase 10.1103/PhysRevB.80.165117} {\bibfield  {journal}
  {\bibinfo  {journal} {Physical Review B}\ }\textbf {\bibinfo {volume} {80}},\
  \bibinfo {pages} {5} (\bibinfo {year} {2009})}\BibitemShut {NoStop}%
\bibitem [{\citenamefont {Holzner}\ \emph {et~al.}(2011)\citenamefont
  {Holzner}, \citenamefont {Weichselbaum}, \citenamefont {McCulloch},
  \citenamefont {Schollw\"{o}ck},\ and\ \citenamefont {von
  Delft}}]{Holzner2011}%
  \BibitemOpen
  \bibfield  {author} {\bibinfo {author} {\bibfnamefont {A.}~\bibnamefont
  {Holzner}}, \bibinfo {author} {\bibfnamefont {A.}~\bibnamefont
  {Weichselbaum}}, \bibinfo {author} {\bibfnamefont {I.~P.}\ \bibnamefont
  {McCulloch}}, \bibinfo {author} {\bibfnamefont {U.}~\bibnamefont
  {Schollw\"{o}ck}}, \ and\ \bibinfo {author} {\bibfnamefont {J.}~\bibnamefont
  {von Delft}},\ }\href {\doibase 10.1103/PhysRevB.83.195115} {\bibfield
  {journal} {\bibinfo  {journal} {Physical Review B}\ }\textbf {\bibinfo
  {volume} {83}},\ \bibinfo {pages} {195115} (\bibinfo {year}
  {2011})}\BibitemShut {NoStop}%
\bibitem [{\citenamefont {Dargel}\ \emph {et~al.}(2012)\citenamefont {Dargel},
  \citenamefont {W\"{o}llert}, \citenamefont {Honecker}, \citenamefont
  {McCulloch}, \citenamefont {Schollw\"{o}ck},\ and\ \citenamefont
  {Pruschke}}]{Dargel2012}%
  \BibitemOpen
  \bibfield  {author} {\bibinfo {author} {\bibfnamefont {P.~E.}\ \bibnamefont
  {Dargel}}, \bibinfo {author} {\bibfnamefont {A.}~\bibnamefont {W\"{o}llert}},
  \bibinfo {author} {\bibfnamefont {A.}~\bibnamefont {Honecker}}, \bibinfo
  {author} {\bibfnamefont {I.~P.}\ \bibnamefont {McCulloch}}, \bibinfo {author}
  {\bibfnamefont {U.}~\bibnamefont {Schollw\"{o}ck}}, \ and\ \bibinfo {author}
  {\bibfnamefont {T.}~\bibnamefont {Pruschke}},\ }\href {\doibase
  10.1103/PhysRevB.85.205119} {\bibfield  {journal} {\bibinfo  {journal}
  {Physical Review B}\ }\textbf {\bibinfo {volume} {85}},\ \bibinfo {pages}
  {205119} (\bibinfo {year} {2012})}\BibitemShut {NoStop}%
\bibitem [{\citenamefont {White}\ and\ \citenamefont
  {Affleck}(2008)}]{White2008a}%
  \BibitemOpen
  \bibfield  {author} {\bibinfo {author} {\bibfnamefont {S.~R.}\ \bibnamefont
  {White}}\ and\ \bibinfo {author} {\bibfnamefont {I.}~\bibnamefont
  {Affleck}},\ }\href {http://link.aps.org/doi/10.1103/PhysRevB.77.134437}
  {\bibfield  {journal} {\bibinfo  {journal} {Physical Review B}\ }\textbf
  {\bibinfo {volume} {77}},\ \bibinfo {pages} {134437} (\bibinfo {year}
  {2008})}\BibitemShut {NoStop}%
\bibitem [{\citenamefont {Feynman}(1954)}]{Feynman1954}%
  \BibitemOpen
  \bibfield  {author} {\bibinfo {author} {\bibfnamefont {R.~P.}\ \bibnamefont
  {Feynman}},\ }\href {\doibase 10.1103/PhysRev.94.262} {\bibfield  {journal}
  {\bibinfo  {journal} {Physical Review}\ }\textbf {\bibinfo {volume} {94}},\
  \bibinfo {pages} {262} (\bibinfo {year} {1954})}\BibitemShut {NoStop}%
\bibitem [{\citenamefont {Feynman}\ and\ \citenamefont
  {Cohen}(1956)}]{Feynman1956}%
  \BibitemOpen
  \bibfield  {author} {\bibinfo {author} {\bibfnamefont {R.~P.}\ \bibnamefont
  {Feynman}}\ and\ \bibinfo {author} {\bibfnamefont {M.}~\bibnamefont
  {Cohen}},\ }\href {\doibase 10.1103/PhysRev.102.1189} {\bibfield  {journal}
  {\bibinfo  {journal} {Physical Review}\ }\textbf {\bibinfo {volume} {102}},\
  \bibinfo {pages} {1189} (\bibinfo {year} {1956})}\BibitemShut {NoStop}%
\bibitem [{\citenamefont {Bijl}\ \emph {et~al.}(1941)\citenamefont {Bijl},
  \citenamefont {de~Boer},\ and\ \citenamefont {Michels}}]{Bijl1941}%
  \BibitemOpen
  \bibfield  {author} {\bibinfo {author} {\bibfnamefont {A.}~\bibnamefont
  {Bijl}}, \bibinfo {author} {\bibfnamefont {J.}~\bibnamefont {de~Boer}}, \
  and\ \bibinfo {author} {\bibfnamefont {A.}~\bibnamefont {Michels}},\ }\href
  {\doibase 10.1016/S0031-8914(41)90422-6} {\bibfield  {journal} {\bibinfo
  {journal} {Physica}\ }\textbf {\bibinfo {volume} {8}},\ \bibinfo {pages}
  {655} (\bibinfo {year} {1941})}\BibitemShut {NoStop}%
\bibitem [{\citenamefont {Haegeman}\ \emph {et~al.}(2012)\citenamefont
  {Haegeman}, \citenamefont {Pirvu}, \citenamefont {Weir}, \citenamefont
  {Cirac}, \citenamefont {Osborne}, \citenamefont {Verschelde},\ and\
  \citenamefont {Verstraete}}]{Haegeman2012a}%
  \BibitemOpen
  \bibfield  {author} {\bibinfo {author} {\bibfnamefont {J.}~\bibnamefont
  {Haegeman}}, \bibinfo {author} {\bibfnamefont {B.}~\bibnamefont {Pirvu}},
  \bibinfo {author} {\bibfnamefont {D.~J.}\ \bibnamefont {Weir}}, \bibinfo
  {author} {\bibfnamefont {J.~I.}\ \bibnamefont {Cirac}}, \bibinfo {author}
  {\bibfnamefont {T.~J.}\ \bibnamefont {Osborne}}, \bibinfo {author}
  {\bibfnamefont {H.}~\bibnamefont {Verschelde}}, \ and\ \bibinfo {author}
  {\bibfnamefont {F.}~\bibnamefont {Verstraete}},\ }\href {\doibase
  10.1103/PhysRevB.85.100408} {\bibfield  {journal} {\bibinfo  {journal}
  {Physical Review B}\ }\textbf {\bibinfo {volume} {85}},\ \bibinfo {pages}
  {100408} (\bibinfo {year} {2012})}\BibitemShut {NoStop}%
\bibitem [{\citenamefont {Draxler}\ \emph {et~al.}(2013)\citenamefont
  {Draxler}, \citenamefont {Haegeman}, \citenamefont {Osborne}, \citenamefont
  {Stojevic}, \citenamefont {Vanderstraeten},\ and\ \citenamefont
  {Verstraete}}]{Draxler2013}%
  \BibitemOpen
  \bibfield  {author} {\bibinfo {author} {\bibfnamefont {D.}~\bibnamefont
  {Draxler}}, \bibinfo {author} {\bibfnamefont {J.}~\bibnamefont {Haegeman}},
  \bibinfo {author} {\bibfnamefont {T.~J.}\ \bibnamefont {Osborne}}, \bibinfo
  {author} {\bibfnamefont {V.}~\bibnamefont {Stojevic}}, \bibinfo {author}
  {\bibfnamefont {L.}~\bibnamefont {Vanderstraeten}}, \ and\ \bibinfo {author}
  {\bibfnamefont {F.}~\bibnamefont {Verstraete}},\ }\href {\doibase
  10.1103/PhysRevLett.111.020402} {\bibfield  {journal} {\bibinfo  {journal}
  {Physical Review Letters}\ }\textbf {\bibinfo {volume} {111}},\ \bibinfo
  {pages} {020402} (\bibinfo {year} {2013})}\BibitemShut {NoStop}%
\bibitem [{\citenamefont {Milsted}\ \emph {et~al.}(2013)\citenamefont
  {Milsted}, \citenamefont {Haegeman},\ and\ \citenamefont
  {Osborne}}]{Milsted2013}%
  \BibitemOpen
  \bibfield  {author} {\bibinfo {author} {\bibfnamefont {A.}~\bibnamefont
  {Milsted}}, \bibinfo {author} {\bibfnamefont {J.}~\bibnamefont {Haegeman}}, \
  and\ \bibinfo {author} {\bibfnamefont {T.~J.}\ \bibnamefont {Osborne}},\
  }\href {http://link.aps.org/doi/10.1103/PhysRevD.88.085030} {\bibfield
  {journal} {\bibinfo  {journal} {Physical Review D}\ }\textbf {\bibinfo
  {volume} {88}},\ \bibinfo {pages} {085030} (\bibinfo {year}
  {2013})}\BibitemShut {NoStop}%
\bibitem [{\citenamefont {Buyens}\ \emph {et~al.}(2013)\citenamefont {Buyens},
  \citenamefont {Haegeman}, \citenamefont {{Van Acoleyen}}, \citenamefont
  {Verschelde},\ and\ \citenamefont {Verstraete}}]{Buyens2013}%
  \BibitemOpen
  \bibfield  {author} {\bibinfo {author} {\bibfnamefont {B.}~\bibnamefont
  {Buyens}}, \bibinfo {author} {\bibfnamefont {J.}~\bibnamefont {Haegeman}},
  \bibinfo {author} {\bibfnamefont {K.}~\bibnamefont {{Van Acoleyen}}},
  \bibinfo {author} {\bibfnamefont {H.}~\bibnamefont {Verschelde}}, \ and\
  \bibinfo {author} {\bibfnamefont {F.}~\bibnamefont {Verstraete}},\ }\href
  {http://arxiv.org/abs/1312.6654} {\  (\bibinfo {year} {2013})},\ \Eprint
  {http://arxiv.org/abs/1312.6654} {arXiv:1312.6654} \BibitemShut {NoStop}%
\bibitem [{\citenamefont {Haegeman}\ \emph
  {et~al.}(2013{\natexlab{a}})\citenamefont {Haegeman}, \citenamefont
  {Michalakis}, \citenamefont {Nachtergaele}, \citenamefont {Osborne},
  \citenamefont {Schuch},\ and\ \citenamefont {Verstraete}}]{Haegeman2013a}%
  \BibitemOpen
  \bibfield  {author} {\bibinfo {author} {\bibfnamefont {J.}~\bibnamefont
  {Haegeman}}, \bibinfo {author} {\bibfnamefont {S.}~\bibnamefont
  {Michalakis}}, \bibinfo {author} {\bibfnamefont {B.}~\bibnamefont
  {Nachtergaele}}, \bibinfo {author} {\bibfnamefont {T.~J.}\ \bibnamefont
  {Osborne}}, \bibinfo {author} {\bibfnamefont {N.}~\bibnamefont {Schuch}}, \
  and\ \bibinfo {author} {\bibfnamefont {F.}~\bibnamefont {Verstraete}},\
  }\href {\doibase 10.1103/PhysRevLett.111.080401} {\bibfield  {journal}
  {\bibinfo  {journal} {Physical Review Letters}\ }\textbf {\bibinfo {volume}
  {111}},\ \bibinfo {pages} {080401} (\bibinfo {year}
  {2013}{\natexlab{a}})}\BibitemShut {NoStop}%
\bibitem [{\citenamefont {Shastry}\ and\ \citenamefont
  {Sutherland}(1981)}]{Shastry1981}%
  \BibitemOpen
  \bibfield  {author} {\bibinfo {author} {\bibfnamefont {B.~S.}\ \bibnamefont
  {Shastry}}\ and\ \bibinfo {author} {\bibfnamefont {B.}~\bibnamefont
  {Sutherland}},\ }\href {\doibase 10.1103/PhysRevLett.47.964} {\bibfield
  {journal} {\bibinfo  {journal} {Physical Review Letters}\ }\textbf {\bibinfo
  {volume} {47}},\ \bibinfo {pages} {964} (\bibinfo {year} {1981})}\BibitemShut
  {NoStop}%
\bibitem [{\citenamefont {Affleck}(1990)}]{Affleck1990}%
  \BibitemOpen
  \bibfield  {author} {\bibinfo {author} {\bibfnamefont {I.}~\bibnamefont
  {Affleck}},\ }\href {\doibase 10.1103/PhysRevB.41.6697} {\bibfield  {journal}
  {\bibinfo  {journal} {Physical Review B}\ }\textbf {\bibinfo {volume} {41}},\
  \bibinfo {pages} {6697} (\bibinfo {year} {1990})}\BibitemShut {NoStop}%
\bibitem [{\citenamefont {Affleck}(1991)}]{Affleck1991}%
  \BibitemOpen
  \bibfield  {author} {\bibinfo {author} {\bibfnamefont {I.}~\bibnamefont
  {Affleck}},\ }\href {\doibase 10.1103/PhysRevB.43.3215} {\bibfield  {journal}
  {\bibinfo  {journal} {Physical Review B}\ }\textbf {\bibinfo {volume} {43}},\
  \bibinfo {pages} {3215} (\bibinfo {year} {1991})}\BibitemShut {NoStop}%
\bibitem [{\citenamefont {Tsvelik}(1990)}]{Tsvelik1990}%
  \BibitemOpen
  \bibfield  {author} {\bibinfo {author} {\bibfnamefont {A.}~\bibnamefont
  {Tsvelik}},\ }\href {\doibase 10.1103/PhysRevB.42.10499} {\bibfield
  {journal} {\bibinfo  {journal} {Physical Review B}\ }\textbf {\bibinfo
  {volume} {42}},\ \bibinfo {pages} {10499} (\bibinfo {year}
  {1990})}\BibitemShut {NoStop}%
\bibitem [{\citenamefont {Chitra}\ and\ \citenamefont
  {Giamarchi}(1997)}]{Chitra1997}%
  \BibitemOpen
  \bibfield  {author} {\bibinfo {author} {\bibfnamefont {R.}~\bibnamefont
  {Chitra}}\ and\ \bibinfo {author} {\bibfnamefont {T.}~\bibnamefont
  {Giamarchi}},\ }\href {\doibase 10.1103/PhysRevB.55.5816} {\bibfield
  {journal} {\bibinfo  {journal} {Physical Review B}\ }\textbf {\bibinfo
  {volume} {55}},\ \bibinfo {pages} {5816} (\bibinfo {year}
  {1997})}\BibitemShut {NoStop}%
\bibitem [{\citenamefont {Konik}\ and\ \citenamefont
  {Fendley}(2002)}]{Konik2002}%
  \BibitemOpen
  \bibfield  {author} {\bibinfo {author} {\bibfnamefont {R.}~\bibnamefont
  {Konik}}\ and\ \bibinfo {author} {\bibfnamefont {P.}~\bibnamefont
  {Fendley}},\ }\href {\doibase 10.1103/PhysRevB.66.144416} {\bibfield
  {journal} {\bibinfo  {journal} {Physical Review B}\ }\textbf {\bibinfo
  {volume} {66}},\ \bibinfo {pages} {144416} (\bibinfo {year}
  {2002})}\BibitemShut {NoStop}%
\bibitem [{\citenamefont {Lou}\ \emph {et~al.}(2000)\citenamefont {Lou},
  \citenamefont {Qin}, \citenamefont {Ng}, \citenamefont {Su},\ and\
  \citenamefont {Affleck}}]{Lou2000}%
  \BibitemOpen
  \bibfield  {author} {\bibinfo {author} {\bibfnamefont {J.}~\bibnamefont
  {Lou}}, \bibinfo {author} {\bibfnamefont {S.}~\bibnamefont {Qin}}, \bibinfo
  {author} {\bibfnamefont {T.-K.}\ \bibnamefont {Ng}}, \bibinfo {author}
  {\bibfnamefont {Z.}~\bibnamefont {Su}}, \ and\ \bibinfo {author}
  {\bibfnamefont {I.}~\bibnamefont {Affleck}},\ }\href {\doibase
  10.1103/PhysRevB.62.3786} {\bibfield  {journal} {\bibinfo  {journal}
  {Physical Review B}\ }\textbf {\bibinfo {volume} {62}},\ \bibinfo {pages}
  {3786} (\bibinfo {year} {2000})}\BibitemShut {NoStop}%
\bibitem [{\citenamefont {Okunishi}\ \emph {et~al.}(1999)\citenamefont
  {Okunishi}, \citenamefont {Hieida},\ and\ \citenamefont
  {Akutsu}}]{Okunishi1999}%
  \BibitemOpen
  \bibfield  {author} {\bibinfo {author} {\bibfnamefont {K.}~\bibnamefont
  {Okunishi}}, \bibinfo {author} {\bibfnamefont {Y.}~\bibnamefont {Hieida}}, \
  and\ \bibinfo {author} {\bibfnamefont {Y.}~\bibnamefont {Akutsu}},\ }\href
  {\doibase 10.1103/PhysRevB.59.6806} {\bibfield  {journal} {\bibinfo
  {journal} {Physical Review B}\ }\textbf {\bibinfo {volume} {59}},\ \bibinfo
  {pages} {6806} (\bibinfo {year} {1999})}\BibitemShut {NoStop}%
\bibitem [{\citenamefont {Damle}\ and\ \citenamefont
  {Sachdev}(1998)}]{Damle1998}%
  \BibitemOpen
  \bibfield  {author} {\bibinfo {author} {\bibfnamefont {K.}~\bibnamefont
  {Damle}}\ and\ \bibinfo {author} {\bibfnamefont {S.}~\bibnamefont
  {Sachdev}},\ }\href {\doibase 10.1103/PhysRevB.57.8307} {\bibfield  {journal}
  {\bibinfo  {journal} {Physical Review B}\ }\textbf {\bibinfo {volume} {57}},\
  \bibinfo {pages} {8307} (\bibinfo {year} {1998})}\BibitemShut {NoStop}%
\bibitem [{\citenamefont {Haegeman}\ \emph {et~al.}(2011)\citenamefont
  {Haegeman}, \citenamefont {Cirac}, \citenamefont {Osborne}, \citenamefont
  {Pi\v{z}orn}, \citenamefont {Verschelde},\ and\ \citenamefont
  {Verstraete}}]{Haegeman2011d}%
  \BibitemOpen
  \bibfield  {author} {\bibinfo {author} {\bibfnamefont {J.}~\bibnamefont
  {Haegeman}}, \bibinfo {author} {\bibfnamefont {J.~I.}\ \bibnamefont {Cirac}},
  \bibinfo {author} {\bibfnamefont {T.~J.}\ \bibnamefont {Osborne}}, \bibinfo
  {author} {\bibfnamefont {I.}~\bibnamefont {Pi\v{z}orn}}, \bibinfo {author}
  {\bibfnamefont {H.}~\bibnamefont {Verschelde}}, \ and\ \bibinfo {author}
  {\bibfnamefont {F.}~\bibnamefont {Verstraete}},\ }\href {\doibase
  10.1103/PhysRevLett.107.070601} {\bibfield  {journal} {\bibinfo  {journal}
  {Physical Review Letters}\ }\textbf {\bibinfo {volume} {107}},\ \bibinfo
  {pages} {070601} (\bibinfo {year} {2011})}\BibitemShut {NoStop}%
\bibitem [{\citenamefont {Fannes}\ \emph {et~al.}(1992)\citenamefont {Fannes},
  \citenamefont {Nachtergaele},\ and\ \citenamefont {Werner}}]{Fannes1992}%
  \BibitemOpen
  \bibfield  {author} {\bibinfo {author} {\bibfnamefont {M.}~\bibnamefont
  {Fannes}}, \bibinfo {author} {\bibfnamefont {B.}~\bibnamefont
  {Nachtergaele}}, \ and\ \bibinfo {author} {\bibfnamefont {R.~F.}\
  \bibnamefont {Werner}},\ }\href {\doibase 10.1007/BF02099178} {\bibfield
  {journal} {\bibinfo  {journal} {Communications in Mathematical Physics}\
  }\textbf {\bibinfo {volume} {144}},\ \bibinfo {pages} {443} (\bibinfo {year}
  {1992})}\BibitemShut {NoStop}%
\bibitem [{\citenamefont {Vidal}(2007)}]{Vidal2007b}%
  \BibitemOpen
  \bibfield  {author} {\bibinfo {author} {\bibfnamefont {G.}~\bibnamefont
  {Vidal}},\ }\href {\doibase 10.1103/PhysRevLett.98.070201} {\bibfield
  {journal} {\bibinfo  {journal} {Physical Review Letters}\ }\textbf {\bibinfo
  {volume} {98}},\ \bibinfo {pages} {5} (\bibinfo {year} {2007})}\BibitemShut
  {NoStop}%
\bibitem [{\citenamefont {Haegeman}\ \emph {et~al.}(2014)\citenamefont
  {Haegeman}, \citenamefont {Mari\"{e}n}, \citenamefont {Osborne},\ and\
  \citenamefont {Verstraete}}]{Haegeman2014}%
  \BibitemOpen
  \bibfield  {author} {\bibinfo {author} {\bibfnamefont {J.}~\bibnamefont
  {Haegeman}}, \bibinfo {author} {\bibfnamefont {M.}~\bibnamefont
  {Mari\"{e}n}}, \bibinfo {author} {\bibfnamefont {T.~J.}\ \bibnamefont
  {Osborne}}, \ and\ \bibinfo {author} {\bibfnamefont {F.}~\bibnamefont
  {Verstraete}},\ }\href {\doibase 10.1063/1.4862851} {\bibfield  {journal}
  {\bibinfo  {journal} {Journal of Mathematical Physics}\ }\textbf {\bibinfo
  {volume} {55}},\ \bibinfo {pages} {021902} (\bibinfo {year}
  {2014})}\BibitemShut {NoStop}%
\bibitem [{\citenamefont {Haegeman}\ \emph
  {et~al.}(2013{\natexlab{b}})\citenamefont {Haegeman}, \citenamefont
  {Osborne},\ and\ \citenamefont {Verstraete}}]{Haegeman2013b}%
  \BibitemOpen
  \bibfield  {author} {\bibinfo {author} {\bibfnamefont {J.}~\bibnamefont
  {Haegeman}}, \bibinfo {author} {\bibfnamefont {T.~J.}\ \bibnamefont
  {Osborne}}, \ and\ \bibinfo {author} {\bibfnamefont {F.}~\bibnamefont
  {Verstraete}},\ }\href {\doibase 10.1103/PhysRevB.88.075133} {\bibfield
  {journal} {\bibinfo  {journal} {Physical Review B}\ }\textbf {\bibinfo
  {volume} {88}},\ \bibinfo {pages} {075133} (\bibinfo {year}
  {2013}{\natexlab{b}})}\BibitemShut {NoStop}%
\bibitem [{pre()}]{preparation}%
  \BibitemOpen
  \href@noop {} {\bibinfo  {journal} {In preparation}\ }\BibitemShut {NoStop}%
\bibitem [{Note1()}]{Note1}%
  \BibitemOpen
\bibfield  {journal} {  }\bibinfo {note} {Thanks to the tensor structure of
  MPS, our method can be implemented with a computational complexity scaling as
  $\protect \mathcal {O}(D^3)$ in the bond dimension.}\BibitemShut {Stop}%
\bibitem [{\citenamefont {Haldane}(1983)}]{Haldane1983a}%
  \BibitemOpen
  \bibfield  {author} {\bibinfo {author} {\bibfnamefont {F.~D.~M.}\
  \bibnamefont {Haldane}},\ }\href {\doibase 10.1016/0375-9601(83)90631-X}
  {\bibfield  {journal} {\bibinfo  {journal} {Physics Letters A}\ }\textbf
  {\bibinfo {volume} {93}},\ \bibinfo {pages} {464} (\bibinfo {year}
  {1983})}\BibitemShut {NoStop}%
\bibitem [{\citenamefont {Takahashi}(1989)}]{Takahashi1989}%
  \BibitemOpen
  \bibfield  {author} {\bibinfo {author} {\bibfnamefont {M.}~\bibnamefont
  {Takahashi}},\ }\href {\doibase 10.1103/PhysRevLett.62.2313} {\bibfield
  {journal} {\bibinfo  {journal} {Physical Review Letters}\ }\textbf {\bibinfo
  {volume} {62}},\ \bibinfo {pages} {2313} (\bibinfo {year}
  {1989})}\BibitemShut {NoStop}%
\bibitem [{\citenamefont {Takahashi}(1994)}]{Takahashi1994}%
  \BibitemOpen
  \bibfield  {author} {\bibinfo {author} {\bibfnamefont {M.}~\bibnamefont
  {Takahashi}},\ }\href {\doibase 10.1103/PhysRevB.50.3045} {\bibfield
  {journal} {\bibinfo  {journal} {Physical Review B}\ }\textbf {\bibinfo
  {volume} {50}},\ \bibinfo {pages} {3045} (\bibinfo {year}
  {1994})}\BibitemShut {NoStop}%
\bibitem [{\citenamefont {White}\ and\ \citenamefont {Huse}(1993)}]{White1993}%
  \BibitemOpen
  \bibfield  {author} {\bibinfo {author} {\bibfnamefont {S.~R.}\ \bibnamefont
  {White}}\ and\ \bibinfo {author} {\bibfnamefont {D.}~\bibnamefont {Huse}},\
  }\href {\doibase 10.1103/PhysRevB.48.3844} {\bibfield  {journal} {\bibinfo
  {journal} {Physical Review B}\ }\textbf {\bibinfo {volume} {48}},\ \bibinfo
  {pages} {3844} (\bibinfo {year} {1993})}\BibitemShut {NoStop}%
\bibitem [{\citenamefont {Sorensen}\ and\ \citenamefont
  {Affleck}(1993)}]{Sorensen1993}%
  \BibitemOpen
  \bibfield  {author} {\bibinfo {author} {\bibfnamefont {E.}~\bibnamefont
  {Sorensen}}\ and\ \bibinfo {author} {\bibfnamefont {I.}~\bibnamefont
  {Affleck}},\ }\href {\doibase 10.1103/PhysRevLett.71.1633} {\bibfield
  {journal} {\bibinfo  {journal} {Physical Review Letters}\ }\textbf {\bibinfo
  {volume} {71}},\ \bibinfo {pages} {1633} (\bibinfo {year}
  {1993})}\BibitemShut {NoStop}%
\bibitem [{\citenamefont {Sorensen}\ and\ \citenamefont
  {Affleck}(1994{\natexlab{a}})}]{Sorensen1994}%
  \BibitemOpen
  \bibfield  {author} {\bibinfo {author} {\bibfnamefont {E.}~\bibnamefont
  {Sorensen}}\ and\ \bibinfo {author} {\bibfnamefont {I.}~\bibnamefont
  {Affleck}},\ }\href {\doibase 10.1103/PhysRevB.49.15771} {\bibfield
  {journal} {\bibinfo  {journal} {Physical Review B}\ }\textbf {\bibinfo
  {volume} {49}},\ \bibinfo {pages} {15771} (\bibinfo {year}
  {1994}{\natexlab{a}})}\BibitemShut {NoStop}%
\bibitem [{\citenamefont {Sorensen}\ and\ \citenamefont
  {Affleck}(1994{\natexlab{b}})}]{Sorensen1994a}%
  \BibitemOpen
  \bibfield  {author} {\bibinfo {author} {\bibfnamefont {E.}~\bibnamefont
  {Sorensen}}\ and\ \bibinfo {author} {\bibfnamefont {I.}~\bibnamefont
  {Affleck}},\ }\href {\doibase 10.1103/PhysRevB.49.13235} {\bibfield
  {journal} {\bibinfo  {journal} {Physical Review B}\ }\textbf {\bibinfo
  {volume} {49}},\ \bibinfo {pages} {13235} (\bibinfo {year}
  {1994}{\natexlab{b}})}\BibitemShut {NoStop}%
\bibitem [{\citenamefont {Ueda}\ and\ \citenamefont
  {Kusakabe}(2011)}]{Ueda2011}%
  \BibitemOpen
  \bibfield  {author} {\bibinfo {author} {\bibfnamefont {H.}~\bibnamefont
  {Ueda}}\ and\ \bibinfo {author} {\bibfnamefont {K.}~\bibnamefont
  {Kusakabe}},\ }\href {\doibase 10.1103/PhysRevB.84.054446} {\bibfield
  {journal} {\bibinfo  {journal} {Physical Review B}\ }\textbf {\bibinfo
  {volume} {84}},\ \bibinfo {pages} {054446} (\bibinfo {year}
  {2011})}\BibitemShut {NoStop}%
\bibitem [{\citenamefont {Zamolodchikov}\ and\ \citenamefont
  {Zamolodchikov}(1979)}]{Zamolodchikov1979}%
  \BibitemOpen
  \bibfield  {author} {\bibinfo {author} {\bibfnamefont {A.~B.}\ \bibnamefont
  {Zamolodchikov}}\ and\ \bibinfo {author} {\bibfnamefont {A.~B.}\ \bibnamefont
  {Zamolodchikov}},\ }\href {\doibase 10.1016/0003-4916(79)90391-9} {\bibfield
  {journal} {\bibinfo  {journal} {Annals of Physics}\ }\textbf {\bibinfo
  {volume} {120}},\ \bibinfo {pages} {253} (\bibinfo {year}
  {1979})}\BibitemShut {NoStop}%
\bibitem [{Note2()}]{Note2}%
  \BibitemOpen
  \bibinfo {note} {We refer to the supplemental material for more details on
  the precision of these values.}\BibitemShut {Stop}%
\bibitem [{\citenamefont {Affleck}\ and\ \citenamefont
  {Weston}(1992)}]{Affleck1992}%
  \BibitemOpen
  \bibfield  {author} {\bibinfo {author} {\bibfnamefont {I.}~\bibnamefont
  {Affleck}}\ and\ \bibinfo {author} {\bibfnamefont {R.}~\bibnamefont
  {Weston}},\ }\href {\doibase 10.1103/PhysRevB.45.4667} {\bibfield  {journal}
  {\bibinfo  {journal} {Physical Review B}\ }\textbf {\bibinfo {volume} {45}},\
  \bibinfo {pages} {4667} (\bibinfo {year} {1992})}\BibitemShut {NoStop}%
\bibitem [{\citenamefont {Hohenberg}\ and\ \citenamefont
  {Brinkman}(1974)}]{Hohenberg1974}%
  \BibitemOpen
  \bibfield  {author} {\bibinfo {author} {\bibfnamefont {P.}~\bibnamefont
  {Hohenberg}}\ and\ \bibinfo {author} {\bibfnamefont {W.}~\bibnamefont
  {Brinkman}},\ }\href {\doibase 10.1103/PhysRevB.10.128} {\bibfield  {journal}
  {\bibinfo  {journal} {Physical Review B}\ }\textbf {\bibinfo {volume} {10}},\
  \bibinfo {pages} {128} (\bibinfo {year} {1974})}\BibitemShut {NoStop}%
\bibitem [{\citenamefont {Giamarchi}(2004)}]{Giamarchi2004}%
  \BibitemOpen
  \bibfield  {author} {\bibinfo {author} {\bibfnamefont {T.}~\bibnamefont
  {Giamarchi}},\ }\href@noop {} {\emph {\bibinfo {title} {{Quantum Physics in
  One Dimension}}}}\ (\bibinfo  {publisher} {Oxford University Press},\
  \bibinfo {address} {New York},\ \bibinfo {year} {2004})\BibitemShut {NoStop}%
\bibitem [{\citenamefont {F\'{a}th}(2003)}]{Fath2003}%
  \BibitemOpen
  \bibfield  {author} {\bibinfo {author} {\bibfnamefont {G.}~\bibnamefont
  {F\'{a}th}},\ }\href {\doibase 10.1103/PhysRevB.68.134445} {\bibfield
  {journal} {\bibinfo  {journal} {Physical Review B}\ }\textbf {\bibinfo
  {volume} {68}},\ \bibinfo {pages} {134445} (\bibinfo {year}
  {2003})}\BibitemShut {NoStop}%
\bibitem [{\citenamefont {Affleck}(2005)}]{Affleck2005}%
  \BibitemOpen
  \bibfield  {author} {\bibinfo {author} {\bibfnamefont {I.}~\bibnamefont
  {Affleck}},\ }\href {\doibase 10.1103/PhysRevB.72.132414} {\bibfield
  {journal} {\bibinfo  {journal} {Physical Review B}\ }\textbf {\bibinfo
  {volume} {72}},\ \bibinfo {pages} {132414} (\bibinfo {year}
  {2005})}\BibitemShut {NoStop}%
\bibitem [{\citenamefont {Bethe}(1931)}]{Bethe1931}%
  \BibitemOpen
  \bibfield  {author} {\bibinfo {author} {\bibfnamefont {H.}~\bibnamefont
  {Bethe}},\ }\href {\doibase 10.1007/BF01341708} {\bibfield  {journal}
  {\bibinfo  {journal} {Zeitschrift f\"{u}r Physik}\ }\textbf {\bibinfo
  {volume} {71}},\ \bibinfo {pages} {205} (\bibinfo {year} {1931})}\BibitemShut
  {NoStop}%
\bibitem [{\citenamefont {Korepin}\ \emph {et~al.}(1997)\citenamefont
  {Korepin}, \citenamefont {Bogoliubov},\ and\ \citenamefont
  {Izergin}}]{Korepin1997}%
  \BibitemOpen
  \bibfield  {author} {\bibinfo {author} {\bibfnamefont {V.~E.}\ \bibnamefont
  {Korepin}}, \bibinfo {author} {\bibfnamefont {N.~M.}\ \bibnamefont
  {Bogoliubov}}, \ and\ \bibinfo {author} {\bibfnamefont {A.~G.}\ \bibnamefont
  {Izergin}},\ }\href@noop {} {\emph {\bibinfo {title} {{Quantum Inverse
  Scattering Method and Correlation Functions}}}}\ (\bibinfo  {publisher}
  {Cambridge University Press},\ \bibinfo {address} {Cambridge},\ \bibinfo
  {year} {1997})\BibitemShut {NoStop}%
\bibitem [{Note3()}]{Note3}%
  \BibitemOpen
  \bibinfo {note} {We refer to the supplemental material for more details on
  this procedure.}\BibitemShut {Stop}%
\bibitem [{Note4()}]{Note4}%
  \BibitemOpen
  \bibinfo {note} {In the supplemental material we provide some preliminary
  results on the application to a spin ladder.}\BibitemShut {Stop}%
\bibitem [{\citenamefont {Essler}\ and\ \citenamefont
  {Konik}(2008)}]{Essler2008}%
  \BibitemOpen
  \bibfield  {author} {\bibinfo {author} {\bibfnamefont {F.}~\bibnamefont
  {Essler}}\ and\ \bibinfo {author} {\bibfnamefont {R.}~\bibnamefont {Konik}},\
  }\href {\doibase 10.1103/PhysRevB.78.100403} {\bibfield  {journal} {\bibinfo
  {journal} {Physical Review B}\ }\textbf {\bibinfo {volume} {78}},\ \bibinfo
  {pages} {100403} (\bibinfo {year} {2008})}\BibitemShut {NoStop}%
\bibitem [{Note5()}]{Note5}%
  \BibitemOpen
  \bibinfo {note} {In textbook derivations of the Bethe equations the
  scattering phase is often taken to be only dependent on the difference of the
  two individual momenta, because of Galilean invariance. For the magnetized
  chain this will only be approximately the case. For the sake of simplicity we
  have assumed this to be the case, however, and it remains a subject for
  further study to track the degree and implications of the breaking of
  Galilean invariance.}\BibitemShut {Stop}%
\end{thebibliography}
\end{document}